\documentclass[a4paper,11pt]{article}
\pdfoutput=1 
\usepackage{graphicx}
\usepackage{jheppub} 

\usepackage[T1]{fontenc} 
\usepackage{multirow}

\title{\boldmath Comparison of general-purpose events generators for particle fluence simulation in LHC environment}

\author[1]{A. Oblakowska-Mucha,\note{Corresponding author}}
\author{and T. Szumlak}

\affiliation{Faculty of Physics and Applied Computer Science,\\
AGH University of Science and Technology \\
Al. Mickiewicza 30, 30-059 Krakow, Poland}

\emailAdd{Agnieszka.Mucha@agh.edu.pl}
\emailAdd{Tomasz.Szumlak@agh.edu.pl}

\abstract{At the LHC era, the detector systems are operating at the harsh hadronic environment with the unprecedentedly high particle flux. Position sensitive silicon devices are usually positioned at the innermost regions of the experimental setups and must cope with highly non-uniform radiation fields. At the end of LHC Run II, fluence in silicon trackers reached 10$^{15}$ n$_{eq}$/cm$^2$. Initial simulation studies predict that the maximal fluence for the HL-LHC may be up to two orders of magnitude higher than the one seen in LHC Run I and Run II. In this paper, two general-purpose physics events generators used for simulation of proton-proton collisions at LHC energies: Pythia 8.2 and DPMJET 3, are compared. Fluences obtained using these models, with the latest tuning to the LHC data, in detectors situated close to the point of proton-proton collisions are determined as well.}

\begin{document} 
\maketitle
\flushbottom

\section{Introduction}
\label{sec:intro}
Large Hadron Collider (LHC) at CERN was designed to collide bunches of protons accelerated to centre-of-mass energy, $\sqrt s$, of up to 14 TeV. To accomplish a rich physics programme, covering high precision Standard Model (SM) measurements and searches of new physics, the LHC is able to provide 2808 bunches with more than 10$^{11}$ protons each and collide them with a frequency of 40 MHz (at the nominal conditions). LHC can also supply lead ion beams. During the years 2010-2012 (Run I) and 2015-2018 (Run II) the proton beams collided with $\sqrt s$=6.5-7 TeV and 13 TeV respectively. The Run III data taking period will start in the year 2021 and, thanks to the new injectors, the instantaneous luminosity will be doubled and reach approximately 2.2$\times$10$^{34}$ cm$^{-2}$s$^{-1}$.  The LHC program will be continued collecting 3000 fb$^{-1}$ of data around the year 2035 (during Run IV called High Luminosity LHC, HL-LHC)~\cite{HL}. Run IV will be preceded by the significant upgrade of the accelerator together with the substantial change in all main detector trackers because of the radiation damage. 

The total cross-section for proton-proton interaction at $\sqrt s$=8 TeV is $\sigma_{tot}$ = (96.07$\pm$0.18$ \\ \pm$0.31) mb~\cite{sigma_tot}, out of this 71.5 mb is related to the inelastic processes. Assuming the nominal LHC luminosity to be 10$^{34}$ cm$^{-2}$s$^{-1}$, in each second of running the experiments about 96$\times$10$^7$ proton-proton ($pp$) interactions occur. At the LHC energies, a few hundreds of particles may be produced per one $pp$ interaction with momenta that range over many orders of magnitude. The planned HL-LHC runs will double this number, making the radiation damage the main concern in any detector system.

 At the end of LHC Run II data taking period the inner parts of silicon trackers of the LHCb and ATLAS were irradiated of the fluence that reached 10$^{15}$ n$_{eq}$/cm$^2$~\cite{velo_fluence}. The increase of the luminosity in Run III will cause an increase of the particle fluence as well, reaching annually value  compared to the Run I and II together. When the level of 10$^{16}$ n$_{eq}$/cm$^2$ is reached, the detectors' proper performance will be significantly affected~\cite{rd50}. The precautions, whereas the operating conditions, must be changed or eventually sensors should be replaced, will be taken based on the simulation of the particle fluence and monitoring of the actual state of the detector.
 
The fluence in the LHC environment is highly non-uniform and comes from particles emerging from the high energy proton-proton collisions and also from the interactions of these particles with the sub-detectors, supports and shielding. The radiation field is composed of charged hadrons and leptons, neutrons and photons. The proportion of the different particle types at a given point of the radiation field depends on the distance and the angle with respect to the proton beams and on the detector's material. The only reliable way to estimate the properties of such complex radiation filed is through the Monte Carlo based simulation. Estimation of the particle fluence is therefore necessarily both for the monitoring of the current state of a detector (to anticipate the change in operation) and for the prediction whereas all components would operate until the end of the planned data taking period. Last but not least, simulation of the fluence is indispensable for the verification of the models for the radiation damage models~\cite{rd50}.

The simulation procedure of particle fluence traversing the LHC detectors contains two components: generation of particles (i.e., the final state system) originating from a proton-proton collision at the LHC energies, and the subsequent simulation of particle transport in the detectors' material. The former is elaborated in this document. Two models of the high energy proton-proton collision used in programs for simulation of particle fluence are compared: Pythia 8.2 and DPMJET 3. Samples are generated without any selection criteria which could bias the comparison.
If any differences are observed at the generator lever, they will be naturally propagated to the discrepancies in the particle transport and fluence simulation, causing the inconsistency in the prediction for the future experiments.   

Another method of evaluation of the particle fluence is based on the reconstruction of particle tracks by standard algorithms. This method considers only charged particles and strongly depends on the detector acceptance and experiments' software. Therefore a comparison of such simulation between experiments is not reliable.

\section{Physics models for hadronic interaction }
Proton is not a fundamental particle but is comprised of quarks and gluons (called partons) interacting via the strong and Coulomb force. The type of the interaction and the composition of the final state depend on the proton energy. At the lowest energies ($\sqrt s$ below 1.5 GeV) protons are considered as charged point-like  objects that scatter elastically without any loss of energy. As the energy increases, the inelastic processes prevail, driven by the interaction at the partonic level. With further rise of available energy, the resolution of the projectile parton reveal finer and finer structure of the proton as in Deep Inelastic Scattering (DIS). At the LHC energies the total proton-proton cross-section is dominated by the inelastic proton-proton scattering and the contribution of elastic part is about 25\%.

Kinematics of $pp$ collision is described by four-momentum transfer $Q^2$, that is regarded as a scale at which proton is probed by the parton, and Mandelstam variable $t$ that describes the momentum transfer between partons. In the scattering experiments $Q^2=-t$ and, due to the uncertainty principle, the resolution power of the parton is proportional to $1/\sqrt t$.
With the increase of centre-of-mass energy $\sqrt s$ and the four-momentum transfer $Q^2$, the finer structure of protons is resolved and more gluons and pairs of sea quarks take part in the scattering. Each of these partons carry fraction of protons' momentum, denoted as $x$.
At high energy, gluon density grows much faster than the quark density therefore LHC is essentially a gluon-gluon collider and provides access to the kinematic region never reached before: high $Q^2$ and low $x$.  Due to the high energy and high density of partons, multiparton scattering at LHC occur very often.
 
\begin{figure}[htb]
\centerline{
\includegraphics[width=17cm]{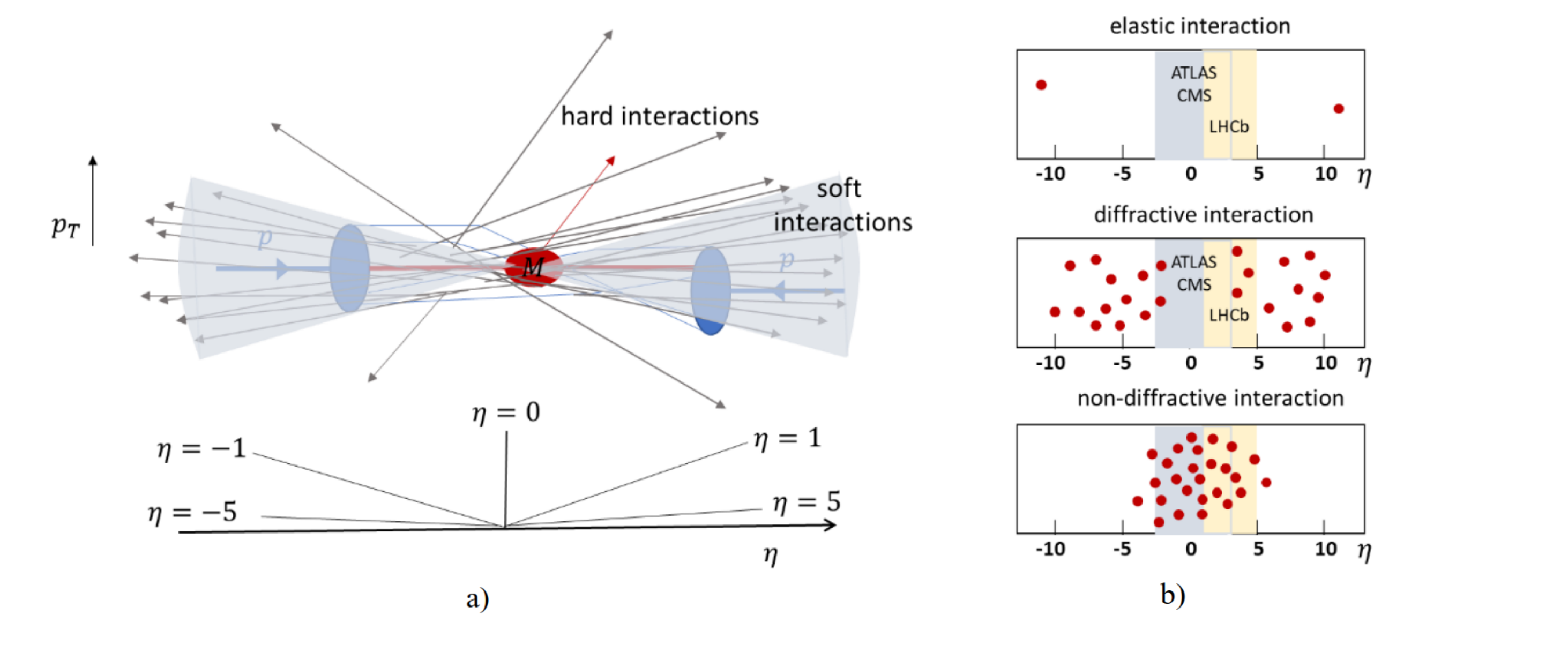}}
\caption{a) Schema of typical $pp$ collision at LHC energy. When scattered parton has large transverse momentum $p_T$ hadron is produced at large polar angle (small pseudorapidity). Such an interaction is called "hard". The majority of interactions are between partons that carry small fraction of proton momentum and give rise to the flux of hadrons (mainly pions) produced at lower angles (high pseudorapidity). The Higgs boson, or any other interesting massive particle $M$, is rather produced in hard interactions, whereas particles originating from soft interactions are the main source of radiation damage in detector closest to the interaction point. b) The range of pseudorapidity for different components of the total $pp$ cross-section with the LHC detectors coverage.}
\label{fig:softandhard}
\end{figure}

 The interaction between protons involve wide spectrum of processes highly dependent on the fraction of the proton's momenta carried by the partons $x$, and the momentum transfer $t$ between partons. Process is usually regarded as "hard" if is characterised by a large momentum component that is transverse to the beam $p_T$, see Fig.~\ref{fig:softandhard}. This is equivalent to the parton interactions with a large $t$, since $t \simeq p_T^2$. However, the $pp$ collisions, even at high energy, are  dominated by soft processes.

Detectors at the LHC experiments are situated around proton collision points, along the beam line (usually considered as the "$z$" axis in the Cartesian coordinate system). The acceptance of the ATLAS and CMS tracking detectors covers the pseudorapidity region\footnote{Pseudorapidity is defined as: $\eta = -ln(tg ~\theta/2)$ and reflects the distribution of the particles polar angle with respect to the beam axis.} $|\eta|<2.5$ (central part), whereas LHCb is capable to track particles at lower angles:  $2<\eta<5$ (forward). Hard processes are therefore reconstructed in the central part of the detectors, whereas most of the soft interaction result in the final states that are produced in the forward direction. 
The schema of a LHC $pp$ event and the pseudorapidity intervals in the LHC experiments are shown in Fig.~\ref{fig:softandhard}.

\subsection{Hard and soft regime in proton-proton cross-section}
The interaction among hadrons consist of the mix of processes that are driven by a few interactions at the partonic level. The hard (or "semi-hard") interactions involve parton-parton interactions with large momentum transfer that is sufficient to resolve the structure of a hadron.    
Hard interactions are driven by processes with high momentum transfer $p_T >$2 GeV/c and can be described by perturbative QCD (pQCD). Hard scattering results in jets in the final state.
Soft interactions are characterized by an energy scale of an order of the hadron size  (1~fm), what correspond to the momentum transfer $|t|\propto 1/R^2 \approx$ 1 (GeV/c)$^2$. At such large distances perturbative QCD cannot be applied. In such events particles in the final state have large longitudinal, but small transverse momentum.

Each hard interaction between partons from two protons is usually accompanied by the soft interaction between the beam remnants (so-called \textit{Underlying Event}, UE). Radiation of gluons in the initial or final state also give rise to the particles with low $p_T$. Therefore one can assume that $pp$ collisions are dominated by soft processes and the description of the proton interactions involve both perturbative and non-perturbative components, the latter being the subject of modelling and introduces considerable uncertainties. Modelling of the hadronic collisions includes the description of the initial state partons density, process of interaction (creation of parton showers) and hadronisation (formation of the stable hadrons). 

 It is hardly possible to establish one, exclusive and exhaustive theory for the description of proton-proton interaction at both hard and soft regions. 
Proton is not only composite but also extended object, which transverse dimension is Lorentz contracted when it is accelerated to the LHC energies. Thus, accelerated protons look like flat disks traveling in a direction perpendicular to the disk surfaces and, due to the time dilatation, quarks and gluons in the same hadron do not interact among themselves (partons are "frozen" as in  Color Glass Condensate model~\cite{CGC}). Therefore, there is neither no interference between the initial state interaction, nor final state interaction in the same hadron. Thus, we can assume that the interaction occurs only between partons from different protons and happens in a very short time during collision. Hence it can be considered as probing of the internal structure of the proton bound state with the resolution depending on the momentum transfer between scattered partons. The probability of a hard scattering with a large momentum transfer depends on the chance of finding two partons, one from each proton, within the short distance. At the LHC energies it is quite high and, occasionally, even more than one hard scattering may occur in addition to soft interactions.
This phenomenon is referred as double parton-scattering (DPS) and is considered in the current physics models for the LHC particle production.  

Thanks to the lack of the interference in the scattering of partons from different hadrons, the description of the inelastic proton-proton collision is based on the concept of factorization of the cross-section~\cite{fact_collins}. It is assumed that the cross-section can be computed as a convolution product of perturbative hard scattering cross-section of point-like partons and universal factors called parton density functions (PDF) that are assumed to be long-distance non-perturbative quantities and are conveniently written in terms of kinematic variables $x$ and $Q^2$: 
\begin{equation}
\label{eq:sigma_tot}
\sigma = \sum_{i,j,k} \int dx_1 dx_2 ~dt ~\sigma^k_{ij} \times f_i^1(x_1,Q^2) f_j^2(x_2,Q^2),
\end{equation}
where $\sigma^k_{ij}$ is the QCD hard-scattering cross-section for the $k$th process between parton  $i$ and $j$, with a momentum transfer $t$. Functions $f_i^{1,2}(x_{1,2},Q^2)$ describe the probability for finding a parton $i$ with a fraction $x$ of the proton momentum ($0<x \leq 1$) when the proton is probed at the scale of $Q^2$. The main problem in the total hadron cross-section is that it can be neither calculated not measured for the full range of $x$ and $Q^2$. One need to know the distribution of partons in a hadron and partonic cross-section, both for very small and maximal momentum transfers.

 Therefore, the realisation of the factorization idea in generators depends on the physics model and requires fixing the momentum scale for hard and soft interaction. 
MC event generators are used to model the final state particle systems produced in the collision and to obtain predictions for the parameters that cannot be determined experimentally, especially in low-p$_T$  region. Phenomenological models for soft hadron interaction used in general-purpose event generator like PYTHIA, SHERPA of HERWIG takes as a start the perturbative parton-parton interaction with the models of parton showers, hadronisation and multiparton (MPI) interaction. This approach enables tuning of the parameters for the better agreement with data and, thanks to the well-known cross-section in the hard regime, provides the predictive power for the LHC physics. It however does not describe well the very soft region. 

Another approach is undertaken by the PHOJET, EPOS or DPMJET generators. These tools use Regge theory, when the interaction between partons comes from the color chains  which hadronize and produce particles. MPIs are introduced by the exchange of additional objects, like hard pomeron. Higher multiplicities in Run II LHC data enforced a general revision of this approach~\cite{generators}. 

\subsection{Proton-proton cross-section}

The total $pp$ cross-section $\sigma_{tot}$ includes elastic $\sigma_{el}$ and inelastic $\sigma_{in}$ processes. The latter contains several components: diffractive $\sigma_{diff}$ (single, double diffractive and central production) and non-diffractive $\sigma_{ND}$ processes:
\begin{equation}
\label{eq:sigma_const}
\sigma_{tot}= \sigma_{el} + \sigma_{in} =  \sigma_{el} + \sigma_{diff} + \sigma_{ND}
\end{equation}

In elastic collision protons do not change the initial energy and move almost along the beam line after the interaction, see Fig~\ref{fig:softandhard}. Elastic interactions are driven by the long-ranged Coulomb interaction and, much stronger, short-ranged strong interaction. The former is well-known and predictable within the QED, the latter is hard to describe since the perturbative QCD is not applicable in the low-momentum-transfer region, where most of the elastic hadron interaction occur.

Inelastic processes are driven by parton-parton interaction described in Eq.~\ref{eq:sigma_tot} and contain both soft (like diffractive) and hard components. 
$\sigma_{tot}$ is not calculable in the framework of the perturbative QCD. Hence the Regge theory, which considers  exchange of set particles that belong to the Regge trajectory as a cause of interaction, is used for the prediction of the $\sigma_{tot}$ high energy behaviour. 

The soft contribution is parametrised by the soft pomeron exchange  which leads to the peripherial production of partons and, eventually, hadrons. In hard interactions, mainly among valence quarks, all intermediate partons are characterised by large virtualities and all scattering amplitudes can be calculated using the pQCD. Sea quarks and gluons share small fraction of proton's energy and interact via non-perturbative small momentum transfer reactions that appear as long cascades of partons. Created sea quark-antiquark pairs are connected by soft pomeron and undergo both soft and hard interactions~\cite{gribov_regge}. In pQCD pomeron is regarded as series of gluon ladders.

 In single diffractive (SD) processes one of the protons survives the collision while the other breaks up and produces some particles in a limited pseudorapidity region. The intact proton moves at very small angle with respect to proton beam line. The SD events are characterised by the large rapidity gap between the outgoing proton and the produced system. In double diffractive (DD) events both protons dissociate into system of particles which are separated by the rapidity gaps. As a special case of diffractive  reaction central production (CD) may be considered. In these processes both protons survive the collision losing only small amount of energy. A few particles (most often two) are produced centrally, with a rapidity gap from each of the proton. CD process are much less frequent from the SD and DD but can provide valuable information of the proton's structure and dynamics. In non-diffractive events a large number of particles are produced but, opposite to the diffractive events, no gap in rapidity occurs. 

The relative contribution of different terms is established based on physical models with experimental inputs. In general, it can be evaluated, that at the LHC energies the elastic cross-section amounts to about 25\% of the total cross-section, non-diffractive part is about 60\% and diffractive interactions about 15\%. These values vary depending on the applied model and $\sqrt s$.

\begin{figure}[htb]
\centerline{
\includegraphics[width=16cm]{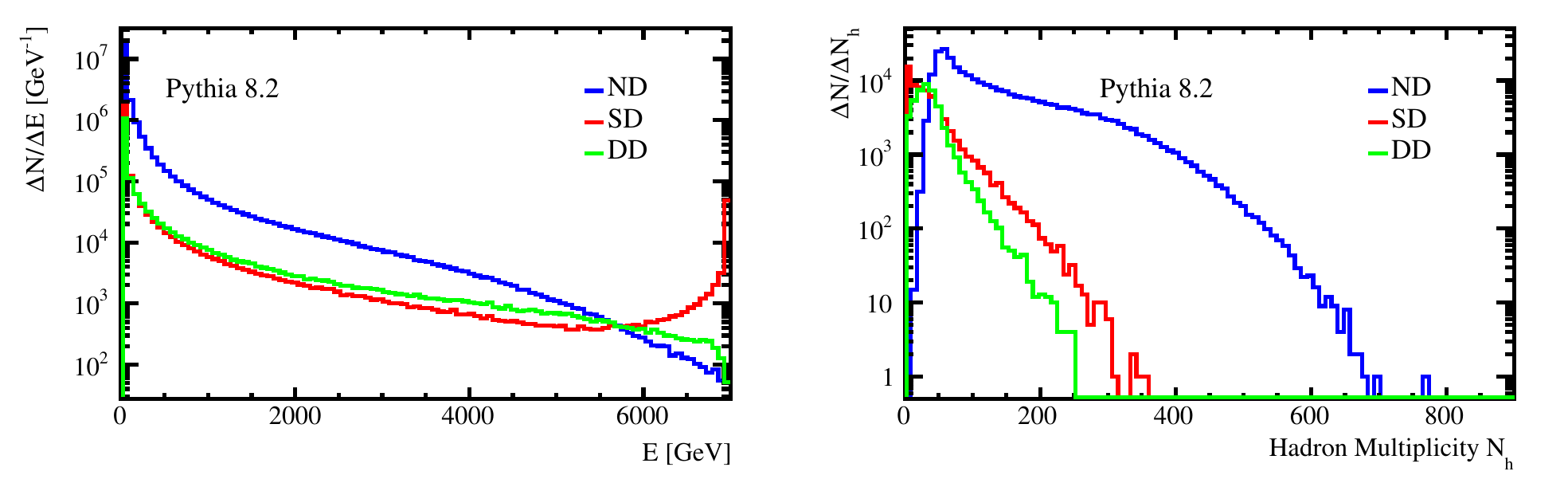} }
\caption{Left: Energy spectrum, right: distribution of particle multiplicity for events generated in non-diffractive (ND), single diffractive (SD) and double-diffractive (DD) processes in 0.5$\times$10$^6$ proton-proton interactions.  }   
\label{fig:en_ND_DD}
\end{figure}

The properties of various final states that contribute to the total proton-proton cross-section differ significantly, as indicated in Fig~\ref{fig:softandhard}. There are two protons with the energy that is equal the beam energy outgoing almost parallel to the beam in elastic interaction, in the SD events a several particles (mostly pions) are produced except one proton at very small polar angle, DD processes provides multiplicities of the order few dozens, whereas most of particles are produced in the detector acceptance through the inelastic ND processes.  The particles' energy spectrum and the distribution of the event multiplicities in different type of proton-proton interactions at $\sqrt s$=14 TeV are shown in Fig.~\ref{fig:en_ND_DD}.

\begin{figure}[htb]
\centerline{
\includegraphics[width=16cm]{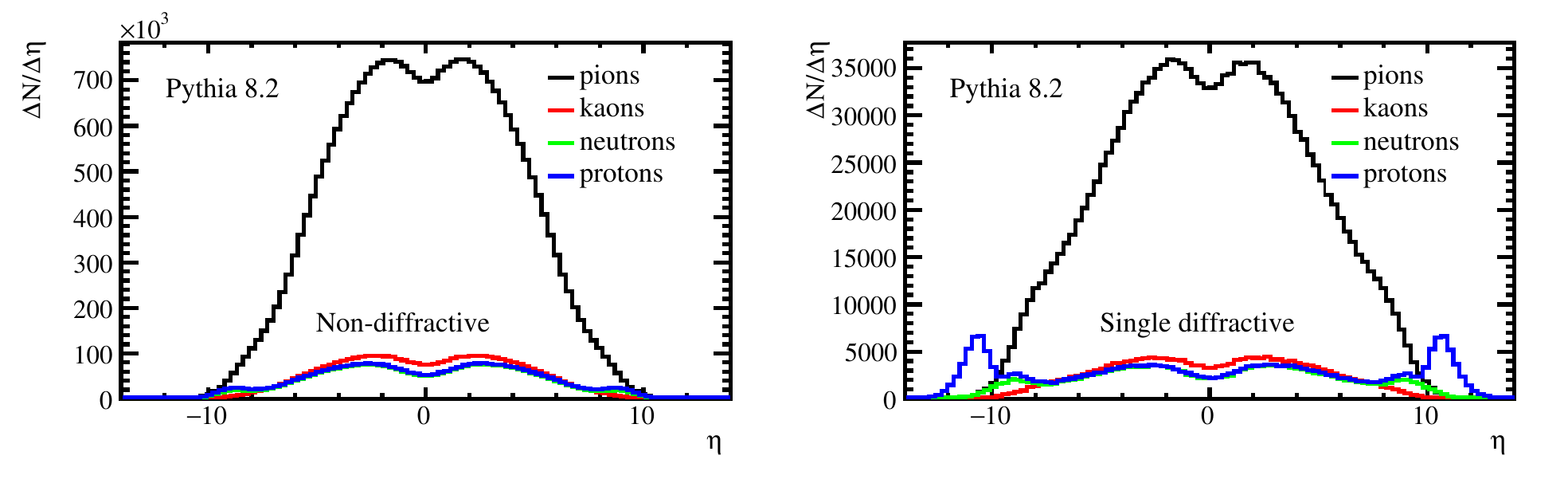} }
\caption{Distribution of the pseudorapidity of the particles participating in: single diffraction (left) non-diffractive inelastic interactions (right). More than 70\% of produced hadrons are pions, 40\% of particles are produced in the central region and 25\% (35\% of total number of protons and 20\% of pions) at $|\eta|>5$ (along the beam line).}   
\label{fig:ND_DD}
\end{figure}

The various components of the total $pp$ cross-section have also a different pseudorapidity distributions, see Fig.~\ref{fig:ND_DD}. These plots show particles produced in half milion proton-proton collision at $\sqrt s$=14 TeV, generated by Pythia 8.2 generator.

 It can be noted that most of the particles reconstructed from $pp$ collision at the LHC energies originate from the inelastic ND processes or are the products of protons dissociation in diffractive events. Since the central collisions are more likely to contain high-p$_T$ partons, most of the particles with higher transverse momentum are produced in the central region of the detectors. 

\subsection{Pythia 8.2}

Pythia is a Monte Carlo (MC) event generator of hadron-hadron collision based on perturbative QCD which describe properly parton scattering above some $p_T$ value ($p_T \gtrsim 2$ GeV/c), extended to the low-$p_T$ region with a models for the soft part.  It generates a wide spectrum of processes, starting from both soft- and hard-QCD processes to Higgs production and exotic particle production and covers all components of the total hadronic cross-section.
This generator keeps on evolving and the currently used version Pythia 8.2 includes tuning originating from the latest LHC results from Run II data taking period ~\cite{Pythia8}. 

The total proton-proton cross-section is described in frame of Donnachie-Landshoff parametrisation as~\cite{Donn_Lann}:  
\begin{equation}
\label{eq:sigma_totpp}
\sigma_{tot}^{pp}(s)= As^P+Bs^{-R}
\end{equation}

All topologies represented by relation~\ref{eq:sigma_const} are included with the amplitudes within Regge theory with one pomeron ($P$) and one Reggeon ($R$) term. The parameters $A,B,P$ and $R$ obtained from the Regge fits to data and give: $A=21.70$, $P=0.0808$, $B=56.08$, $R=0.4525$~\cite{Pythia8}.
The prediction of the the $\sigma_{tot}$ behaviour as the $\sqrt s$ increases are revised as soon as new experimental results are available. One of the major correction had to be applied after the observation from the first runs of LHC at higher $\sqrt s$,  that the parton-parton cross section became larger than the total proton-proton cross-section. It indicated that more than one interaction among the partons occurred in one $pp$ collision. Multiple Parton Interaction (MPI) phenomenon was expected and therefore, based on experimental evidences, generators had to be corrected for the better agreement.

The elastic cross-section is related to the total cross-section via the optical theorem~\cite{Coll_Regge} and is parametrised as an exponentially decreasing function of the Mandelstam invariant $t$. The inelastic component is obtained by the subtraction of the elastic cross-section from the total cross-section: $\sigma_{in}^{pp}(s)=\sigma_{tot}^{pp}(s)-\sigma_{el}^{pp}(s)$. The split of the inelastic cross-section into SD, DD and CD components is based on the specific Pythia parametrisations which are available for possible tuning. Eventually, the non-diffractive part of hadronic interaction (often called in experiment \textit{minimum bias event}) is evaluated by integrating the diffractive components and subtracting them from $\sigma_{in}^{pp}$.

Recent results from TOTEM, ALPHA and LHCb updated information on $\sigma_{tot}^{pp}(s)$ and $\sigma_{el}^{pp}(s)$ at LHC energies $\sqrt s$ =7-8 TeV and 13 TeV~\cite{MPI}. It resulted in more reliable parametrisation of the Eq.~\ref{eq:sigma_totpp} and revision of the previous version of the generators. The models for diffractive interactions were populated with hard-pomeron exchange with the additional possibility of MPIs and initial and final state radiation as well~\cite{MPI}.

The limitation of the Eq.~\ref{eq:sigma_tot} comes from the substantial contribution of the soft, non-perturbative component of the $pp$ interaction. The cross-section above arbitrary chosen $p_{Tmin}$ provides the divergence as $p_T \rightarrow 0$ ($p_T$ is the transverse momentum of the outgoing parton in the parton-parton center-of-mass frame): 
\begin{equation}
\label{eq:sigma_ptmin}
\sigma(p_{Tmin})= \int^{s/4}_{p_{Tmin}} dp_T^2 \frac{d\sigma}{dp_T^2}.
\end{equation}
The scale $Q_{min}^2=p_{Tmin}^2$ can be regarded as a point that connects the soft and hard components of the generator. The choice of $p_{Tmin}$ is of a crutial meaning - the lower it is chosen the higher average number of MPIs and particles in the final state. 
The perturbative parton-parton scattering cross-section given by Eq.~\ref{eq:sigma_tot} and \ref{eq:sigma_ptmin}, dominated by the $t$-channel gluon exchange, is divergent at low-momentum transfers as $1/t^2\sim 1/p_T^4$,  The divergence in Eq.~\ref{eq:sigma_ptmin}  may be regularised by the introduction of the threshold parameter $p_{T0}$  as: $1/p_T^4 \rightarrow 1/(p_T^2+p_{T0}^2)^2$. 
In such a way the perturbative limit is reached as $p_T\rightarrow 0$ with a smooth fall at a scale $p_{T0}$~\cite{MPI}. This approach is supported by the fact that at low $Q^2$ (low $p_T$) the resolution of the probing projectile is poor and the partons in proton are screened by one another. Therefore as $p_T \rightarrow 0$ the cross-section takes small but non-zero value that depends on the matter distribution inside the hadron~\cite{minbias_pred}.

A dependency on the proton-proton centre-of-mass energy should be introduced in addition since at higher energies partons are probed at smaller $x$, where the parton density increases and the distance of colour screening decreases:
$p_{T0}=p_{T0}^{ref}(\sqrt s/\sqrt s_0)^\epsilon $. 
The parameter $\sqrt s_0$ is given at a reference energy and $p_{T0}^{ref}$ is  $p_{T0}$ at $\sqrt s_0$. It means that at given $\sqrt s$  the number of MPIs depends on the $p_{T0}$, smaller values of $p_{T0}$ result in more MPIs because of the increase of the MPI cross-section and higher value of the average particle multiplicity. In that way the parameter $p_{T0}$ is scaled with the respect to reference value $p_{T0}^{ref}$ that is tuned to the experimental results at given $s_0$. The introduction of the $p_{T0}$ in Pythia generator allowed to include the soft scattering regions into hard scattering regime with the price of a special parameter tuning.

The strong couple constant $\alpha_s$ become also divergent at low $p_T$ and had to be regulated by the cut-off parameter $p_{T0}$ as well. It resulted in change of the scale of $\alpha_s$ since it had to be determined in the scale of  $\alpha_s(p_T^2+p_{T0}^2)$ instead of $\alpha_s(p_{T}^2)$. 

Quite a lot parton distribution functions (PDFs) are recorded in the LHAPDF libraries~\cite{lhapdf} and available in Pythia 8.2. Initially PDFs included a leading-order contribution only but the recent event tunes provided by ATLAS and CMS included next-to-leading-order (NLO). The parameters were tuned to the latest LHC results and improved the charged multiplicity distribution from both, low- and high- side, especially for the central rapidity values.

To summarize: the production of particles in proton-proton collisions simulated by Pythia 8.2 generator depends on a number of parameters. The most significant are: $p_{T0}^{ref}$, $\epsilon$, $\alpha_s$. The parameter $p_{T0}$ separates the perturbative from the non-perturbative regions and depends on the centre-of-mass energy $\sqrt s$. Therefore, the reference value $p_{T0}^{ref}$ at chosen reference value $\sqrt{s_0}$  is set with a power-like dependency on $\sqrt s$ with parameter $\epsilon$ .

\begin{figure}[htb]
\centerline{
\includegraphics[width=16cm]{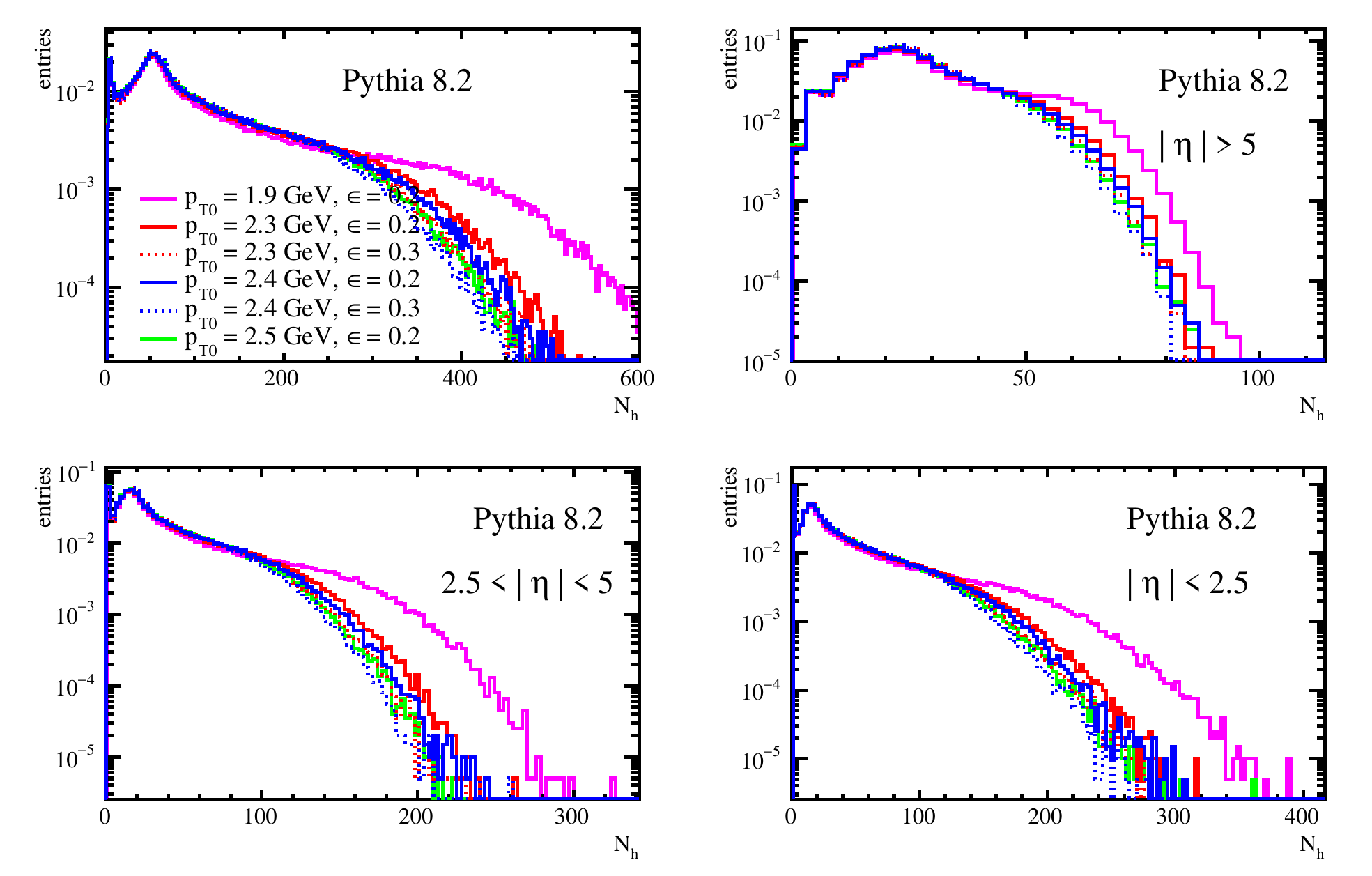} }
\caption{Distribution of hadron multiplicities for different settings of Pythia 8.2 parameters $p_{T0}^{ref}$ and $\epsilon$. Multiplicity is calculated for: full polar angle, $|\eta|$>5, 2.5<$|\eta|$<5, and $|\eta|$<2.5. }   
\label{fig:mult_pythia}
\end{figure}

The comparison of the hadron (pions, protons, kaons, neutrons) multiplicity generated with different settings of $p_{T0}$ and $\epsilon$ is shown in Fig.~\ref{fig:mult_pythia}. The values of parameters were chosen according to the tuning provided by the LHC experiments after Run I~\cite{MPI}. The value $\alpha_s$ was set to $\alpha_s$=0.130 at $\sqrt s_0$=7 TeV according to the experimental tuning. 

All distributions were generated at the proton-proton centre-of-mass-energy $\sqrt s$=14 TeV. Three, important for the experiments, regions in pseudorapidity were defined: $|\eta|$>5 - particles are produced very close to the beam line, beyond the acceptance of any experiment, 2.5<$|\eta|$<5  - in the forward direction, within the LHCb acceptance and $|\eta|$<2.5 - the central detectors.

Smaller values of $p_{T0}$ result in higher number of particles produced and higher hadron multiplicities. The power $\epsilon$ has less influence on the multiplicity distributions, albeit higher value increases the $p_{T0}$ cut-off thus reduces the number of particles from collision.  

In one proton-proton collision at $\sqrt s$=14 TeV about 100 hadrons (charged pions, kaons, protons and neutrons) are produced.
The mean multiplicity varies between 81-105 for the $p_{T0}^{ref}$=2.0-2.5 GeV/c. The difference in settings of the $\epsilon$ parameter from the range $\epsilon$=0.2-0.3 (for the same $p_{T0}^{ref}$) are at the level of 5\%. The distributions of hadron multiplicities generated with different settings are depicted in Fig.~\ref{fig:mult_pythia}. This analysis showed, that one set that result in much higher multiplicities (above 110):  $p_{T0}^{ref}$= 1.9 GeV and $\epsilon$=0.2. This is displayed for better understanding on the tendency: if the cut-off of hard scale was set as a low value, the number of produced particles turned out to increase substantially. The fist estimation based on predictions before the LHC era assumed the $p_{T0}^{ref}$ to be slightly above 1.2 and provided more than 150 hadrons in one proton-proton collision at $\sqrt s$=14 TeV. The results from LHC Run I and first part of Run II shifted the parameters towards higher values and events were simulated with lower multiplicities.     

\subsection{DPMJET 3}

The concept of the DPMJET event generator is based on two-component Dual Parton Model (DPM) which consist of soft hadronic processes described by the exchange of pomerons within the Regge theory~\cite{Coll_Regge}, and hard processes described by perturbative parton scattering~\cite{DPM}. Description of the hadronic final states is provided by both non-perturbative and perturbative QCD included in one common framework and separated by the parameter $p_{Tcuttoff}$. 
DPMJET inherited the model for the proton-proton interaction from the PHOJET generator~\cite{DPMJET3}. 

The scattering of protons in DPM is described by the topologically complicated diagrams with multiple exchanges of pomerons in the $t$-channel. The dominant contribution includes elementary soft interaction between protons which each are split into coloured systems of valence quark and diquark. Two  chains are stretched between quark and diquark from different protons forming color neutral systems that  eventually move apart and hadronize. 
As energy of protons increases, the diagrams with topological expansion of the chains between sea partons give rise to multiparton scattering.
In hadron-hadron interactions several pomerons can be exchanged in parallel, what stands for MPIs.  This concept comprises exchange of a single, double, triple or even loop of pomerons, hard pomerons and soft pomerons. Any of experimentally known process, like single or double diffraction, can be attributed to one of the types of exchanged pomerons, or combination of them. 

Differential elastic proton-proton cross-section  is evaluated as a sum of scattering amplitudes expanded as a series of partial waves with different angular momentum.
The total cross section is related to the elastic cross-section via optical theorem, therefore the inelastic cross-section $\sigma_{in}$ in DPMJET 3 generator is an internal element of the model and includes diffractive processes without any external tuning. The model can be changed 
as soon as one can obtain the appropriate measurements of $\sigma_{tot}$ or $\sigma_{el}$. 

The DPMJET 3 is the latest version embedded in the FLUKA package ~\cite{FLUKA,FLUKA_rev}. The current revision of the program, bundled with FLUKA, was revised and updated to face up with the comparison with LHC data at $\sqrt s$=7 and 13 TeV and differs significantly from the standalone, not tuned, version of the program~\cite{DPMJET3_rev}.
The revised model includes one soft and hard pomerons together with a "effective" reggeon. Therefore, the same model is used for soft and hard components, with fixed, but tuned to data, parameters. This differentiates DPMJET 3 from Pythia, where a vast of parameters can be changed according to the experiments' preferences and different purposes.

\section{Comparison of generators for particle production}

Both generators, DPMJET 3 and Pythia, use factorisation theorem, which allows hadronic cross-sections to be expressed in terms of parton-level cross-sections convoluted with hadron partonic densities. DPMJET 3 originally contained a description of the soft interaction only, and the hard QCD processes were included at the later stage. Pythia, on the contrary, was intentionally dedicated to model the high energy collisions with high transverse momentum transfers, and the size of the soft component was tuned to follow the experimental results. 

Both generators have the same model of hadronisation and use PDF parametrisations from the same LHAPDF library~\cite{lhapdf}. 

Although physics models employed in the Pythia 8.2 and DPMJET 3 generators differ in the approach to the soft and hard components of the total cross-section, both are tuned to the latest experimental data and should not provide substantially different results. Regardless of the total $pp$ cross-section is the same, the contribution of diffractive events may be different, leading to different angular distributions. 

Radiation damage in the bulk of silicon sensors depends on the number of particles produced in the collision, the angle with respect of the proton beamline and the energy spectrum of particles traversing the detector. Therefore, in this study, the comparison of event multiplicities, particles transverse momenta, angular distribution and energy spectra are provided. 

\subsection{Multiplicity distribution and energy spectra}

Depending on the version of the program and parameters used, the number of particles created in one proton-proton collision may differ significantly. Protons, neutrons, pions and kaons, which are regarded as \textit{hadrons} in this document, are particles which are of the crucial meaning for the radiation damage in bulk of silicon. 

The average multiplicity of hadrons produced in both generators is shown in Table~\ref{tab:mult_DP}. This table also contains the composition of hadron flux given with respect to the number of hadrons produced in one interaction. Elastic proton-proton collisions are excluded from this comparison. 
\begin{table}[htb]
    \centering
\begin{tabular}{c|c|c|c|c|c}
            &  $N_h/pp$ &   $N_\pi/N_h$ [\%]   & $N_p/N_h$ [\%] & $N_n/N_h$ [\%] & $N_K/N_h$  [\%] \\ \hline \hline
Pythia 8.2	&   81-105	           & 73.6	        &  8.8	         &  8.2	    &  9.4 \\
DPMJET 3	&   78	               & 82.7	        &  6.3	         &  4.5	    &  6.5
\end{tabular}
    \caption{The average number of hadrons $N_h/pp$ produced in one proton-proton collision and fraction of pions, protons, neutrons and kaon calculated with respect to total number of hadrons. In case of Pythia 8.2 values are calculated for $p_{T0}^{ref}$ from the interval 2.0-2.5 GeV/c. For DPMJET 3 the default FLUKA 2011.2x.6 settings are considered \cite{FLUKA, FLUKA_rev}. }
    \label{tab:mult_DP}
\end{table}

The average number of hadrons produced in one proton-proton collision is slightly higher (5\% for $p_{T0}^{ref}$=2.3 GeV/c)  in the case of Pythia 8.2 with respect to DPMJET 3. This effect reflects the origin of generators - the family of Pythia generators were created to describe the high-multiplicity high-p$_T$ events with the extension to the soft processes. Probabilities of protons and neutrons production are similar due to isospin invariance; kaons are more often produced because of their lower mass. DPMJET 3 produces a higher number of pions than Pythia 8.2, which is understandable in soft processes, whereas the production of heavier baryons requires more momentum transfer and occurs less often. 

\begin{figure}[htb]
\centerline{
\includegraphics[width=16cm]{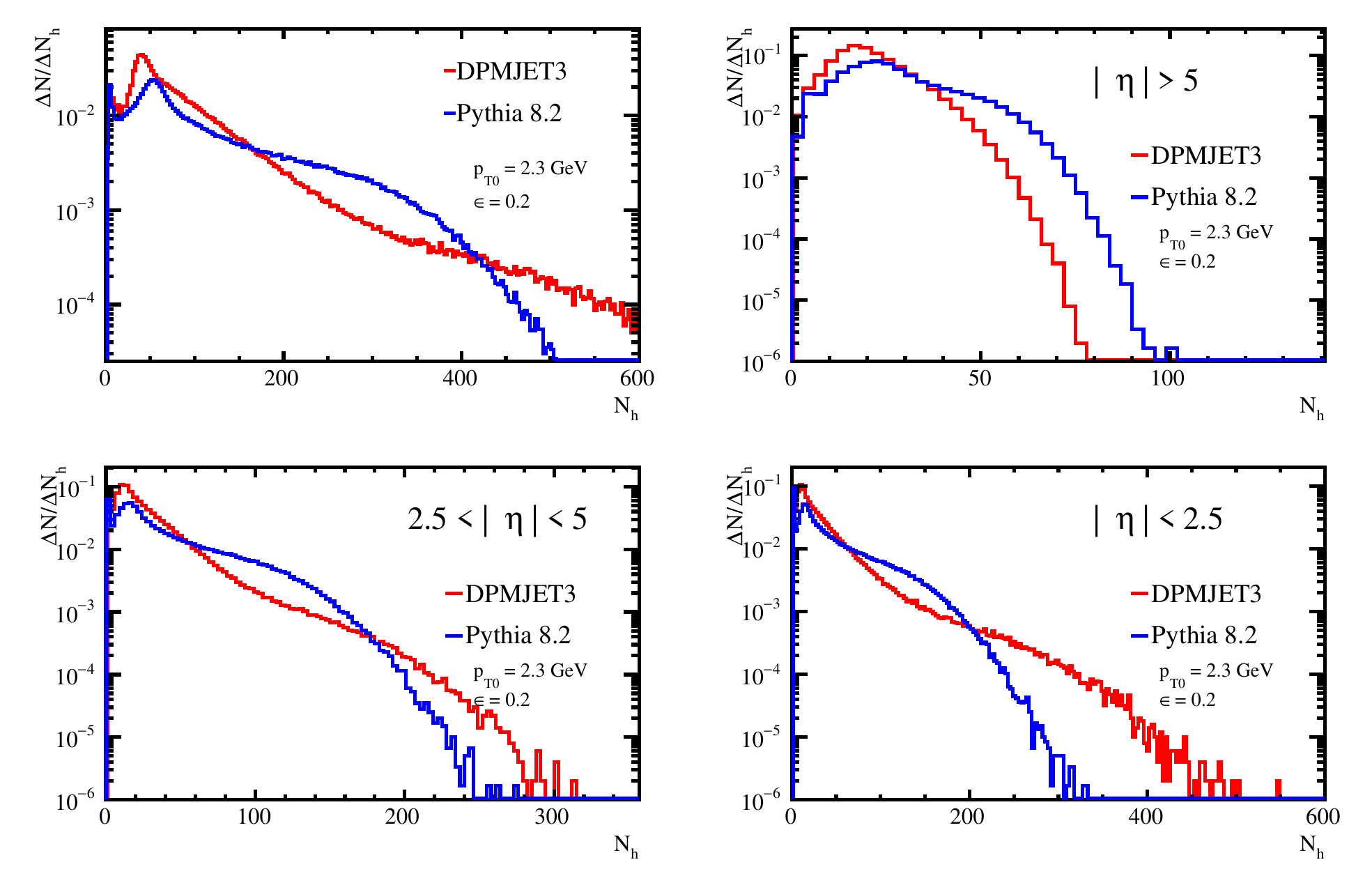} }
\caption{Distribution of multiplicity, i.e. number of hadrons (pions, protons, neutrons, kaons) produced in one proton-proton collision in Pythia 8.2 with $p_{T0}^{ref}$=2.3 GeV/c and DPMJET 3. Multiplicity is calculated for: full polar angle, $|\eta|$>5, 2.5<$|\eta|$<5, and $|\eta|$<2.5. }  
\label{fig:mult_DP}
\end{figure}

Distribution of the multiplicity for one of Pythia settings ($p_{T0}^{ref}$=2.3 GeV/c, $\epsilon$=0.2) and DPMJET 3 is depicted in Fig.~\ref{fig:mult_DP}. It shows that the choice of parameters affects the number of produced particles considerably. Multiplicity is also given in three experimentally important regions of pseudorapidity: particles produced in $|\eta|$>5 that are too close to the beamline to be detected, the region $2.5<|\eta|<5$  is considered as the forward production, whereas $|\eta|<2.5$  represents a typical central detector, see Fig.~\ref{fig:mult_DP}. 

DPMJET 3 generator generates more low-multiplicity events than Pythia 8.2 and very few events with more than 400 hadrons per interaction. The majority of particles are produced in the central region because partons in proton collisions may either propagate as beam remnants moving in the forward direction or populate the central regions with a high number of low-p$_T$ particles. 

\begin{figure}[htb]
\centerline{
\includegraphics[width=16cm]{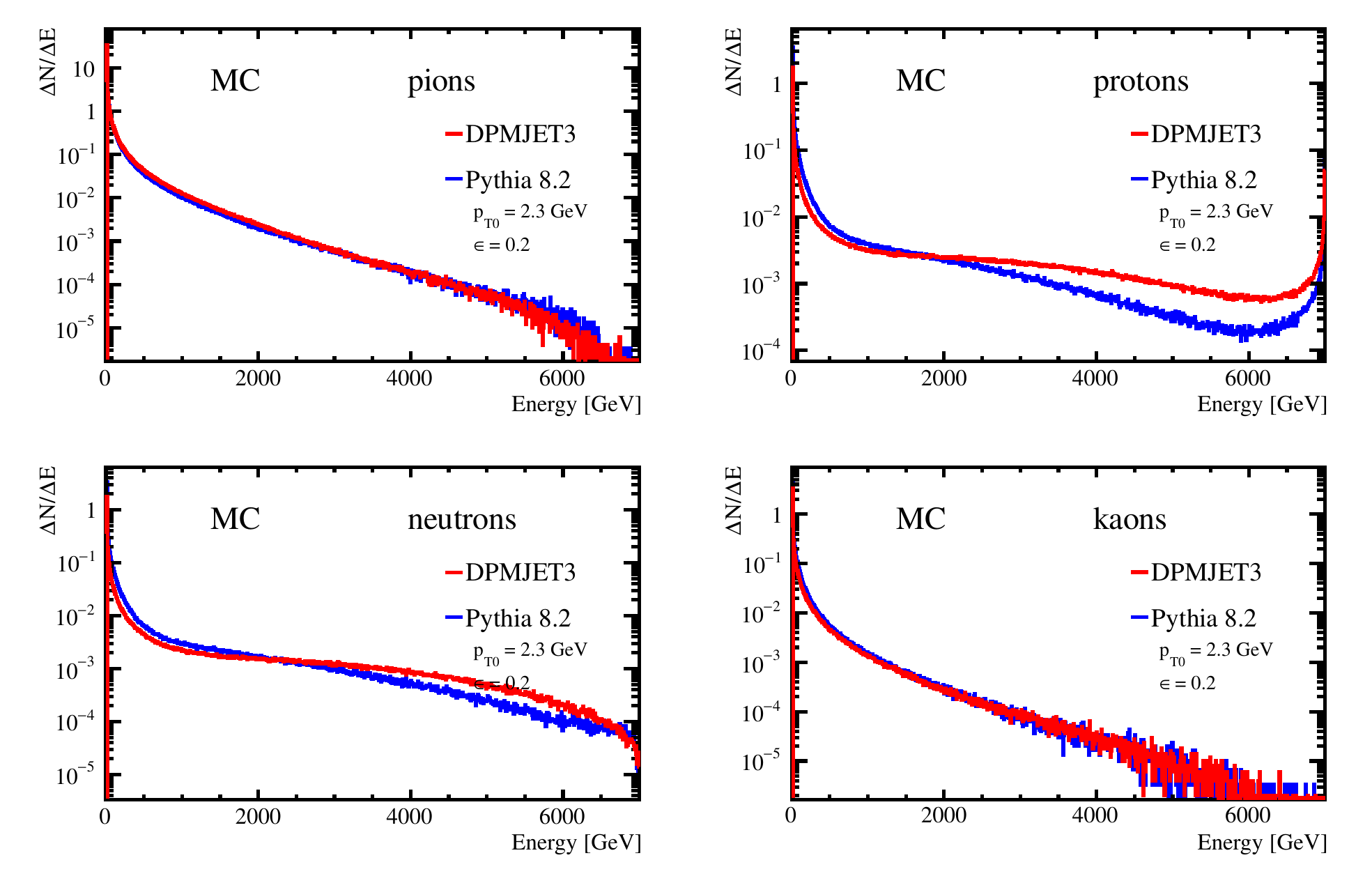} }
\caption{Energy spectrum of pions, protons, neutrons and kaone produced in proton-proton collisions at $\sqrt s$=14 TeV by Pythia 8.2 and DPMJET 3 generators.  Distributions show spectrum of particles produced in one $pp$ collision (logarithmic scale).  }  
\label{fig:energy_DP}
\end{figure}

The energy spectra plotted separately for each type of particles, are shown in Fig.~\ref{fig:energy_DP}. Although both generators produce a similar number of pions with comparable spectra shapes, apparently the number of protons with the highest energies (from diffractive events) produced by DPMJET 3 exceeds the Pythia 8.2 case. Observed differences in the energy spectrum are small and caused only by the differences in the total number of particles. This effect is understandable since parameters of the models in both generators are tuned to the LHC data and samples correspond to the same centre-of-mass energy $\sqrt s$=14 TeV. 

\begin{figure}[htb]
\centerline{
\includegraphics[width=16cm]{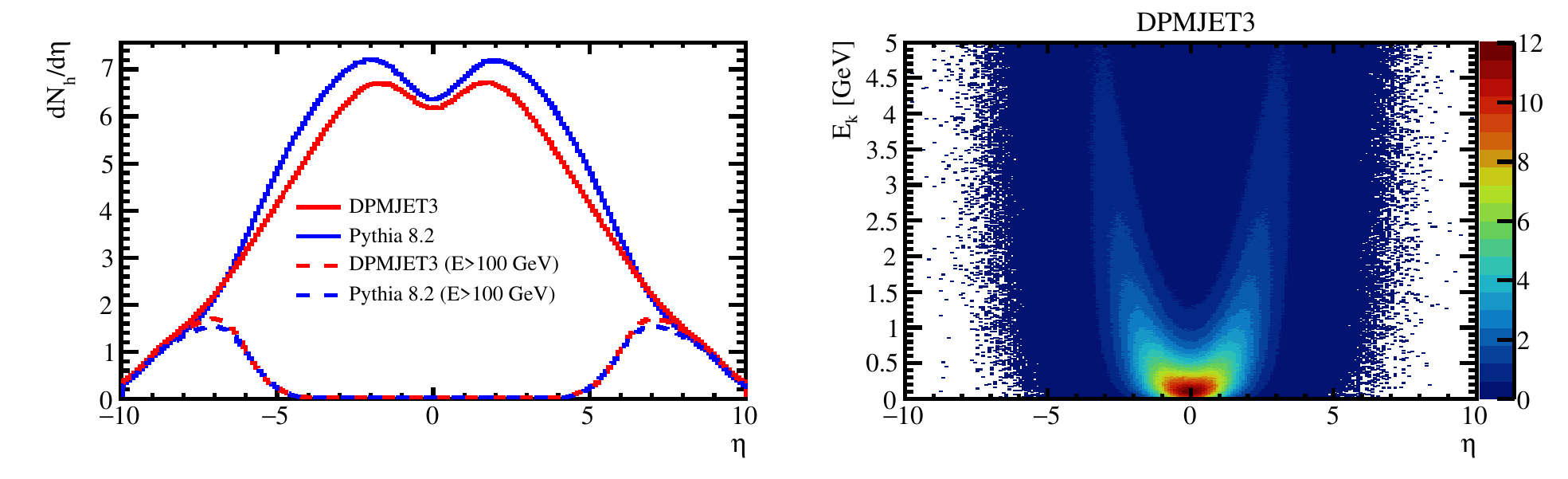} }
\caption{Left: distribution of the pseudorapidity of particles produced in Pythia 8.2 and DPMJET 3 generators.  Particles with energy greater than 100 GeV have pseudorapidity $|\eta|$>4, therefore are beyond the acceptance of the LHC tracking detectors. Distributions are scaled to one proton-proton collision. Right: dependency of the kinetic energy on pseudorapidity (DPMJET 3). Most of the energy is carried by particles with energy below 3 GeV in the central region of the pseudorapidity.  }   
\label{fig:eta_en}
\end{figure}

Distribution of pseudorapidity for all particles with a subset of particles with the energy above 100 GeV\footnote{Tracks reconstructed with good quality usually have momentum below 100 GeV/c. Therefore particles with energy above 100 GeV may not be reconstructed by standard tracking algorithms but still contribute to the fluence and interact with the tracking detectors. } is shown in Fig.~\ref{fig:eta_en}. The majority of particles are produced in the central and forward region, but protons carry the highest energies in the beam pipe region. They originate from the diffractive low-multiplicity events and very rarely can be detected in LHC silicon trackers. 

\begin{figure}[htb]
\centerline{
\includegraphics[width=16cm]{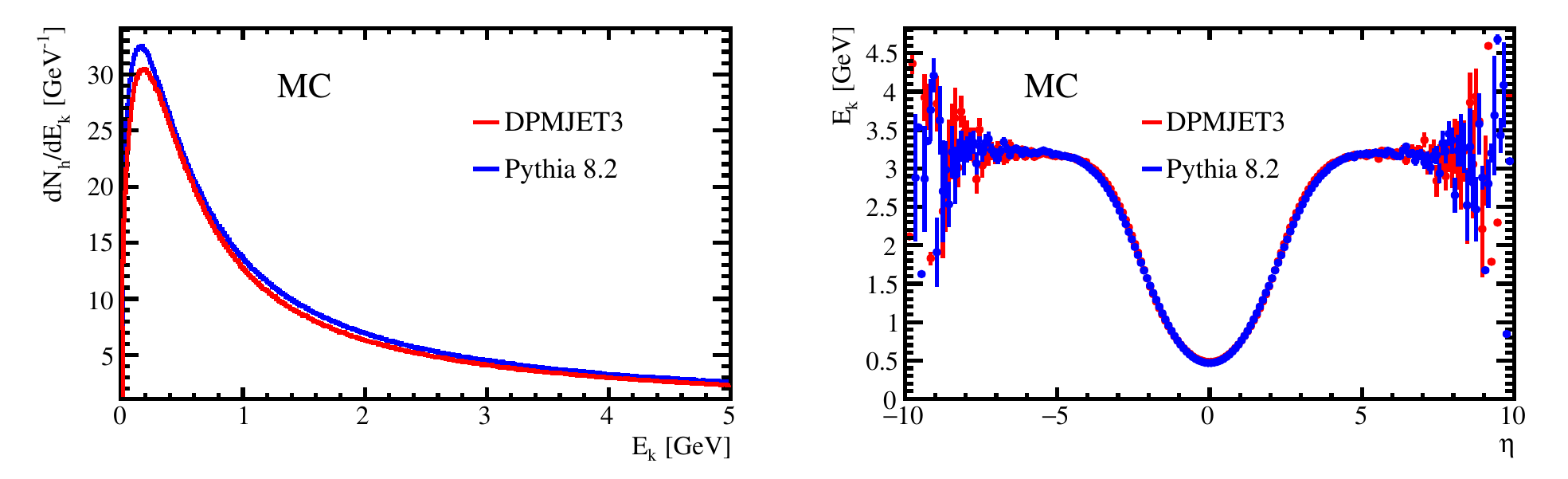} }
\caption{Low-energy spectrum of hadrons produced in one proton-proton collisions at $\sqrt s$=14 TeV by Pythia 8.2 and DPMJET 3 generators (left). Dependency of the kinetic energy on pseudorapidity (right).   }   
\label{fig:lowen}
\end{figure}

Mean energy of particles is about 250 MeV and most of the energy is carried by particles produced in the central values of pseudorapidity, see Fig.~\ref{fig:eta_en} and ~\ref{fig:lowen}.

\subsection{Transverse momenta distribution}

\begin{figure}[htb]
\centerline{
\includegraphics[width=16cm]{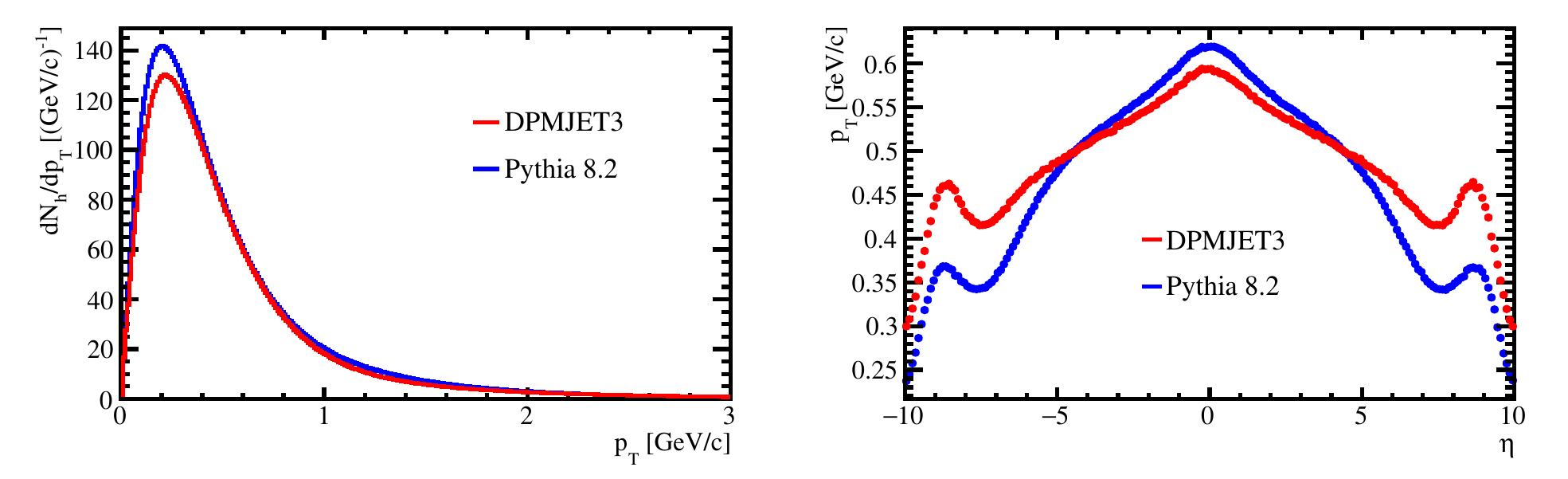} }
\caption{Left: distribution of the transverse momentum p$_T$ of hadrons produced in one proton-proton collisions at $\sqrt s$=14 TeV by Pythia 8.2 and DPMJET3 generators. Right: dependency of the particles' transverse momentum p$_T$ on pseudorapidity $\eta$.   }   
\label{fig:pt}
\end{figure}

Distribution of the particles' transverse momenta $p_T$ shows agreement between both generators with the small excess of DPMJET 3 events with higher $p_T$ which originate from particles with high energy, see the left pane of  Fig.~\ref{fig:pt}. More distinctive difference is visible in the distribution of transverse momenta $p_T$ as a function of pseudorapidity $\eta$ (right plot in Fig.~\ref{fig:pt}). DPMJET 3 generates relatively more low-$p_T$ particles at higher pseudorapidity, i.e. in the very forward region. In consequence, one observes a drop in production of such particles in the central region where the majority of Pythia 8.2 particles are generated.

\begin{figure}[htb]
\centerline{
\includegraphics[width=16cm]{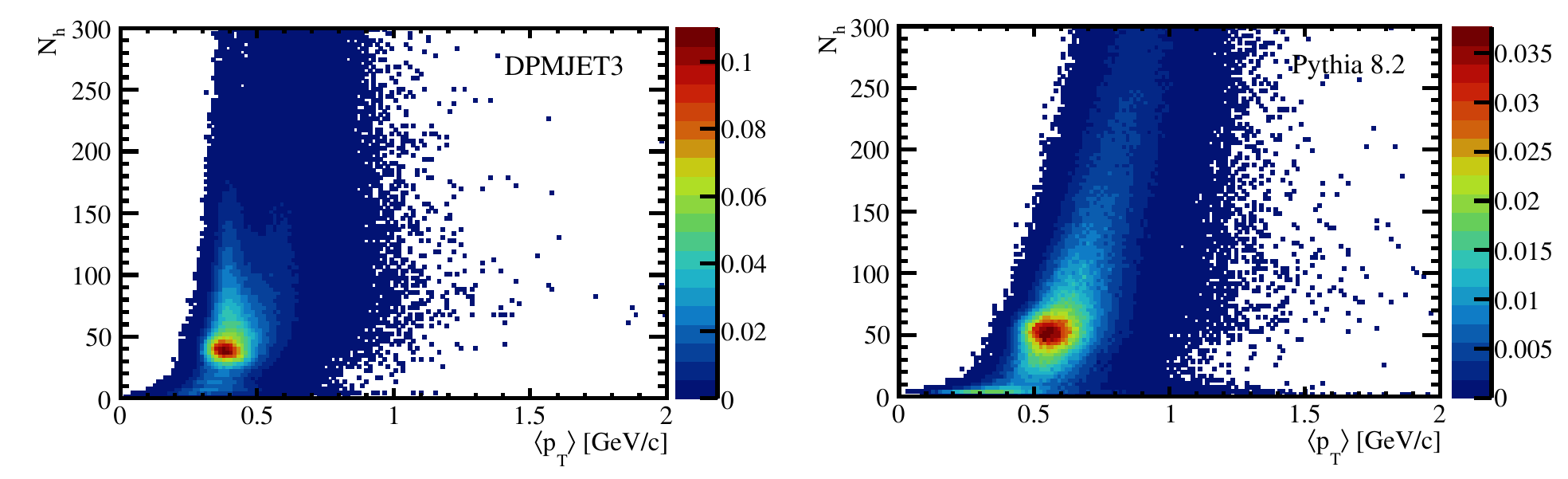} }
\caption{Event multiplicity as a function of mean transverse momentum  $\langle p_T \rangle $ of hadrons in the event:  DPMJET 3 (left), Pythia 8.2 (right).   }  \label{fig:meanpt}
\end{figure}

\begin{figure}[htb]
\centerline{
\includegraphics[width=16cm]{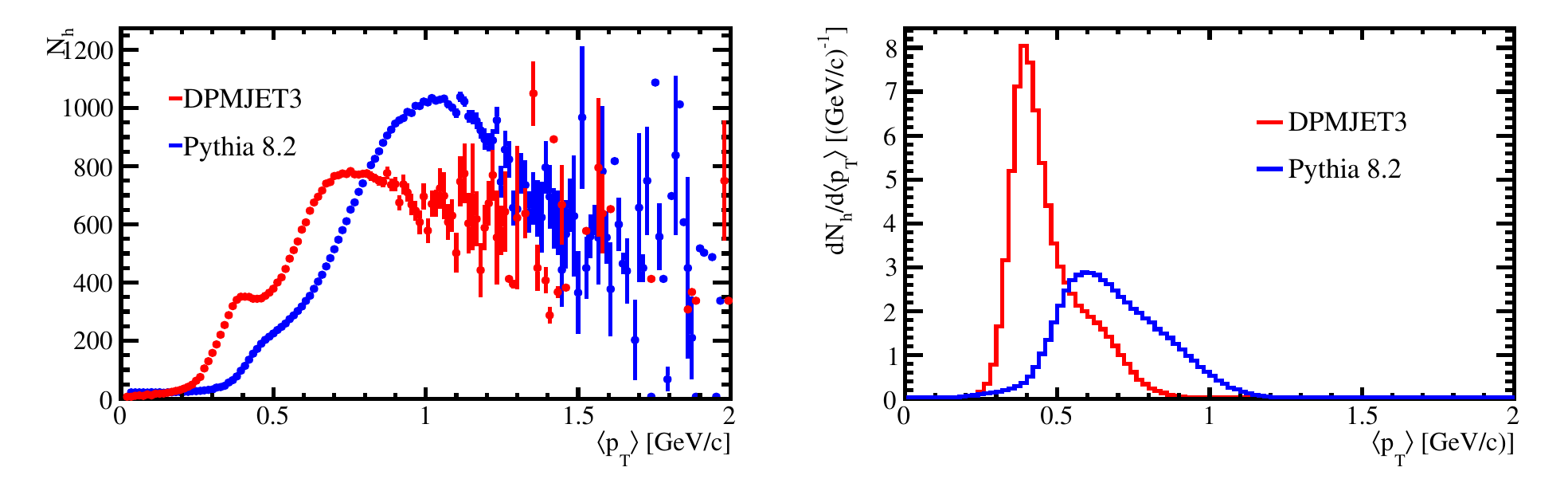} }
\caption{Left: multiplicity of the DPMJET 3 and Pythia 8.2 events as a function of the mean event transverse momentum $\langle p_T \rangle $. Right: comparison of $\langle p_T \rangle$ distribution of events produced in one proton-proton collision between DPMJET 3 and PYTHIA 8.2.   }   
\label{fig:meanpt_proj}
\end{figure}

The correlation of the mean momentum of the produced particles with the corresponding event multiplicity $N_h$ is shown in Fig.~\ref{fig:meanpt}, where the comparison of mean $p_T$ of events generated by DPMJET 3 and Pythia 8.2 is presented. It is noticeable, especially on the projection of these distributions in Fig.~\ref{fig:meanpt_proj}, that Pythia 8.2 events contain more events with higher $\langle p_T \rangle $. 


\subsection{Minimum bias events}

 Minimum-bias data comprise of the events selected with the minimal experimental impact, i.e., without applying any triggers. This class of events is regarded as an experimental realisation of non-diffractive inelastic interactions and is usually used for data-MC comparison. The straightforward experimental selection criteria for each particle are used for comparison: at least one the particle in the event must be within detector acceptance $|\eta|$<5, the kinetic energy must be lower than 100 GeV and the transverse momentum $p_T$ greater than 250 MeV/c.

\begin{figure}[htb]
\centerline{
\includegraphics[width=16cm]{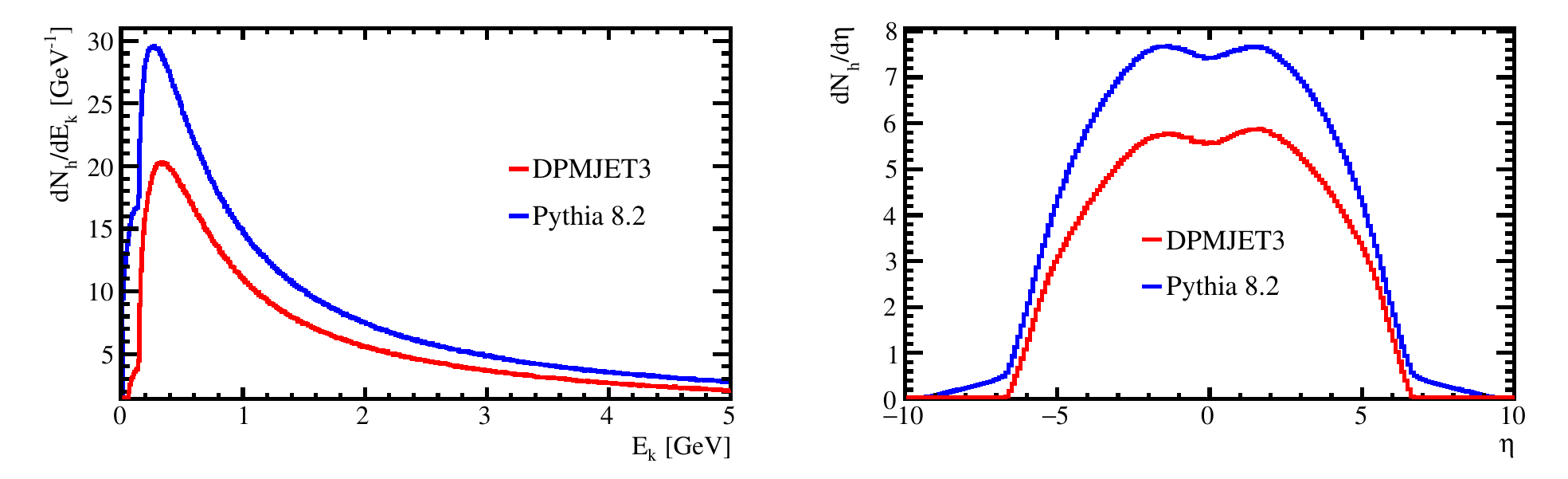} }
\caption{Spectrum of particles' kinetic energy (left) and distribution of their pseudorapidity (right) in one proton-proton interaction for minimum bias events. Events are produced by Pythia 8.2, and DPMJET 3 generators.  }   
\label{fig:en_minbias}
\end{figure}

The spectrum of kinetic energy and pseudorapidity distribution, depicted in Fig.~\ref{fig:en_minbias}, indicate that minimum bias cuts have more severe effects in case of the particles produced by DPMJET 3 generator. In the case of Pythia 8.2, almost 90\% of particles passed these cuts, whereas, in the case of DPMJET 3, this is only 70\%. After applying the cuts, the maximum of $\eta$ distribution in case of events generated by Pythia 8.2 is higher by approximately 20\%  with respect to DPMJET 3, whereas the difference for all generated particles, before the minimum bias selection criteria, is only 4\% (compare Fig.~\ref{fig:eta_en}). 

A more significant rejection rate of particles produced by DPMJET 3 reflects the softer structure of $\langle p_T \rangle$  distribution.  The most substantial effect pertains for pions and protons from diffractive events.

\begin{figure}[htb]
\centerline{
\includegraphics[width=16cm]{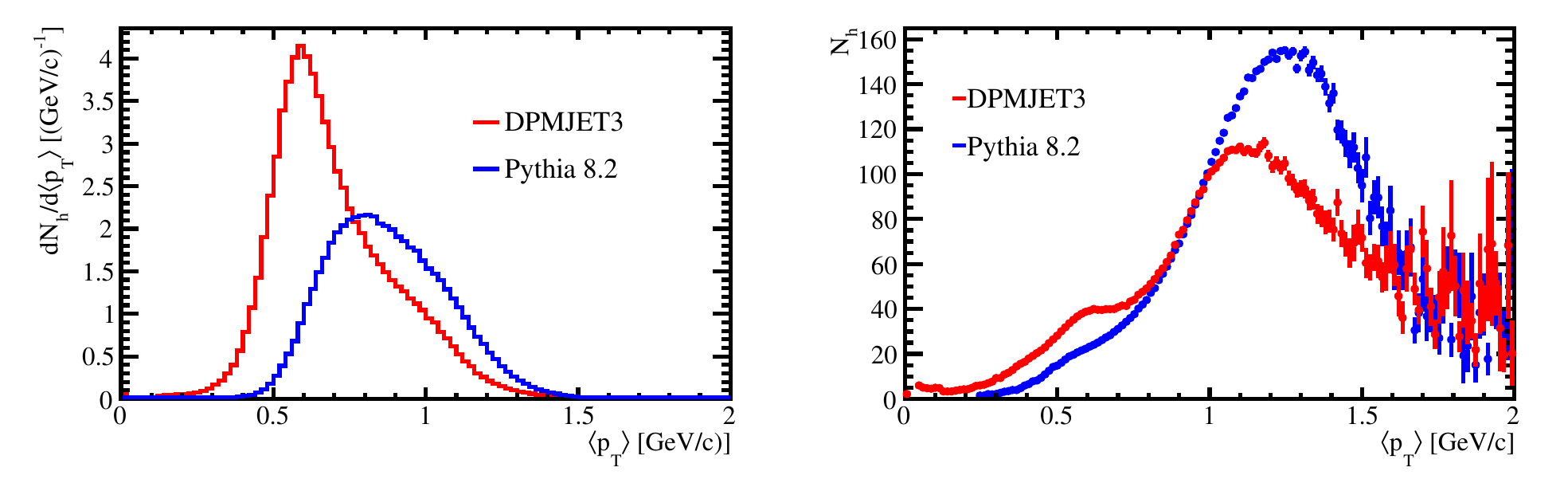} }
\caption{Transverse momentum distribution (left) and hadron multiplicity as a function of $\langle p_T \rangle$ (right) of particles from minimum bias events generated by DPMJET 3 and Pythia 8.2. }   
\label{fig:pt_minbias}
\end{figure}

The distribution of  $\langle p_T \rangle $ and the dependency of the minimum-bias events multiplicity $N_h$ on the  $\langle p_T \rangle $, defined with above criteria, are depicted in Fig.~\ref{fig:pt_minbias}. The comparison illustrates that the Pythia 8.2 generates minimum bias events that are characterised by the harder structure.

The comparison of selected features of particles produced by the respective generators such as energy, momentum, transverse momentum and pseudorapidity has been done. The most interesting result shows about 10\% higher event multiplicity prediction of Pythia 8.2. There are no significant differences in the energy spectrum, but events generated by Pythia 8.2 have higher mean transverse momentum and populates more the central region of pseudorapidity. Very generic experimental criteria imposed on generated events show their influence on pseudorapidity and transverse momentum distribution - in both generators particles produced at the very forward region are not accepted by the minimum bias cuts, but, also, the number of DPMJET 3 particles is further reduced, mainly from the central region. This result shows the potential impact on the calculation of fluence because the minimum bias criteria applied at the generator level might reduce the number of reconstructable particles in case of DPMJET 3.

\section{Fluence simulation in the LHC environment}

The impact of different physics models used for the generation of LHC events may influence the distributions of particle fluence predicted for the detectors, especially at small distances from the interaction point (IP). The impact of secondary particles created in nuclear reactions with the material present within the detector acceptance or stray radiation becomes more important in more distant places of the detector,

For this study, a simulation tool FLUKA is used~\cite{FLUKA, FLUKA_rev}. This is a diverse software platform that includes event generator, a package for geometry description and finally, the particle transport code to simulate interaction with matter. DPMJET 3 generator is embedded in FLUKA and is used as a default generator for the proton-proton collisions. It can be replaced with other generators, like Pythia 8.2, by modifying a specialised interface routine. 

The FLUKA package provides a set of tools for the visualisation and calculation of the variety of dosimetry parameters, like neutron equivalent and particle fluence, total ionising dose, deposited energy and more. Fluence is defined as the number of particles incident on a sphere of given cross-sectional area perpendicular to the direction of each particle. Fluence can also be expressed equivalently, as the lengths of tracks of the particles' trajectories in a unit of volume~\cite{icru}. Thus, in general, the fluence depends on particle multiplicity in an event, energy and polar angle of their tracks with respect to the beamline ($z$-axis). 

The damage occurring to the device is usually normalised to the fluence of neutrons with the kinetic energy of 1 MeV, which would result in the deposition of the same nonionising energy causing equivalent damage to the material. The 1 MeV equivalent neutron fluence $\phi_{eq}$ is calculated internally in the FLUKA program and considers the energy spectrum of the damaging particles and the standard tables with the damage functions~\cite{niel}. 

\subsection{Geometry of the typical LHC experiment}

The physics program of any LHC experiment requires accurate information on the position of production and decay vertices and very efficient track reconstruction. For better precision, the first measured point on track should be as close as possible to the point of protons collision (IP). This is necessary for the life-time measurements and to resolve multiple primary vertices that are created in collision with more than one proton-proton interaction (so-called pile-up events). Therefore, highly granulated silicon detectors are usually situated in the very close proximity to the IP. 

\begin{figure}[htb]
\centerline{
\includegraphics[width=16cm]{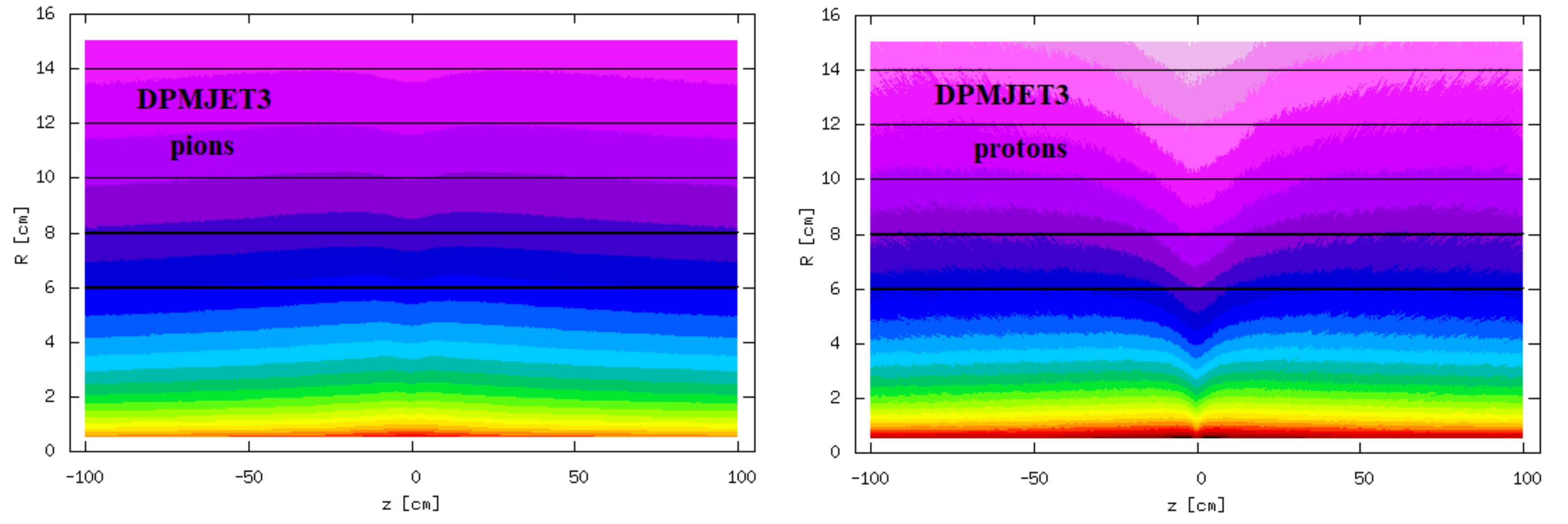} }
\caption{Two-dimensional distribution of pion (left) and proton (right) fluence. Detector consists of five 300 $\mu$m thick silicon cylindrical barrels of radius 6 cm, 8 cm. 10 cm, 12 cm and 14 cm. Simulation is based on a sample with 10$^5$ events from $pp$ interactions at $\sqrt s$=14 TeV generated by DPMJET 3 and reconstructed with the FLUKA program. IP is at the $(x, y, z)$ = $\{0, 0, 0\}$ cm. } 
\label{fig:2D_pr_pi_fluence}
\end{figure}

A typical LHC silicon tracker comprises of the cylindrical barrels with silicon sensors placed parallel to the beamline (along the $z$ axis), see Fig.~\ref{fig:2D_pr_pi_fluence}. For the analysis presented in this paper we created a hypothetical detector with five such layers including 300-$\mu$m silicon sensors. The radii of the barrels are: 6 cm, 8 cm, 10 cm, 12 cm and 14 cm. The collisions of protons at the $\sqrt s$=14 TeV are provided by the DPMJET 3 and Pythia 8.2 generators, the particle transport and track reconstruction is performed within FLUKA package.

\begin{figure}[htb]
\centerline{
\includegraphics[width=16cm]{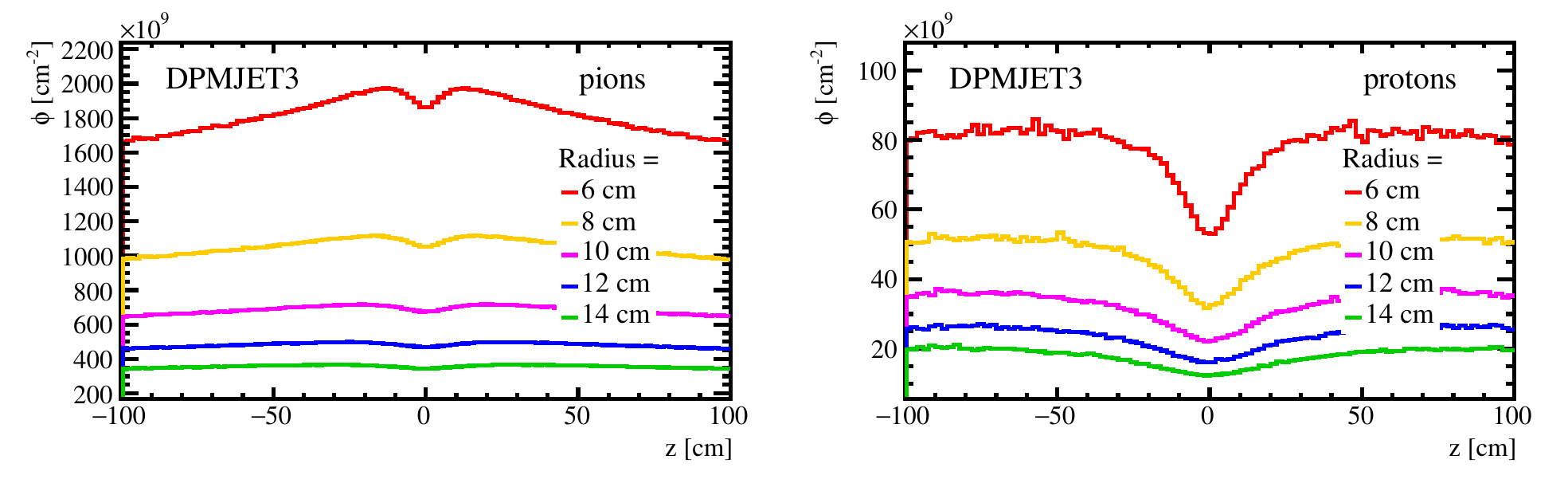} }
\caption{Pion (left) and proton (right) fluence as a function of the distance from the IP. Events are simulated by DPMJET 3 generator and tracks are reconstructed in FLUKA. Plots show the fluence at five radii in five layers of cylindrical barrels (geometry from Fig.~\ref{fig:2D_pr_pi_fluence}). Values are scaled to 1 fb$^{-1}$ ($\sigma_{pp}$= 80 mb).} 
\label{fig:1D_pi_pr_fluence}
\end{figure}

Two-dimensional distribution of pion and proton fluence obtained for this geometry is depicted in Fig.~\ref{fig:2D_pr_pi_fluence}. Dependency of pions and protons fluence on the distance from IP, for the increasing radius of the detector, is shown in Fig.~\ref{fig:1D_pi_pr_fluence}. Fluence is calculated for the amount of data that correspond to 1 fb$^{-1}$ of luminosity assuming the same value for the inelastic cross-section for both generators, $\sigma_{pp}$= 80 mb.

It is evident that in any detector geometry, the highest fluence is expected at small radial distances from the IP because of the highest density of tracks. However, the $z$-dependency has more complex behaviour. It is observed that the particle fluence has a minimum at $z$ = 0 cm since there are hardly any particles produced at large angles to the proton beam direction. This is especially visible in case of protons, which are more often produced at lower angles, whereas pions are more abundant at higher angles. At the high-$z$ tail the distribution of fluence saturates, see Fig.~\ref{fig:1D_pi_pr_fluence}. One can understand it as a purely geometrical effect: when particle traverses consecutive layers at different radii, the track lengths measured in equal detector thickness is the same. The volume element in which tracks' length is measured, increases with the radius, so the tracks' length density decreases for higher radius. But when fluence is evaluated further from the IP in $z$-direction, this effect is compensated by the higher density of particles produced at lower polar angles. 
Therefore, the particle fluence, calculated as the track-length density, is approximately constant at detectors further from the IP than about $z$>40 cm. 

The fluence of pions, protons, neutrons and kaons is usually expressed in the standard neutron equivalence units ($\phi_{eq}$), considering the number of particles, their energy and polar angle. Particle fluence distributions, for respective particle types, are weighted with the energy-dependent displacement damaged functions $D(E)$ and combined~\cite{niel}. The damage function accounts for both the cross-section for displacing silicon atoms and the energy released in creating displacements. Most of the particles produced in the collision have energy below 3 GeV. Damage functions are experimentally determined curves that weight the damage caused by a different type of particles with respect to a neutron of 1 MeV kinetic energy~\cite{niel}. In the case of particles with kinetic energies above 1 GeV, this function is constant 
but for lower energies is higher and depends strongly on energy and varies significantly between different types of particles.
Therefore, low energy particles contribute the most significant to the neutron equivalence fluence. 

\begin{figure}[htb]
\centerline{
\includegraphics[width=16cm]{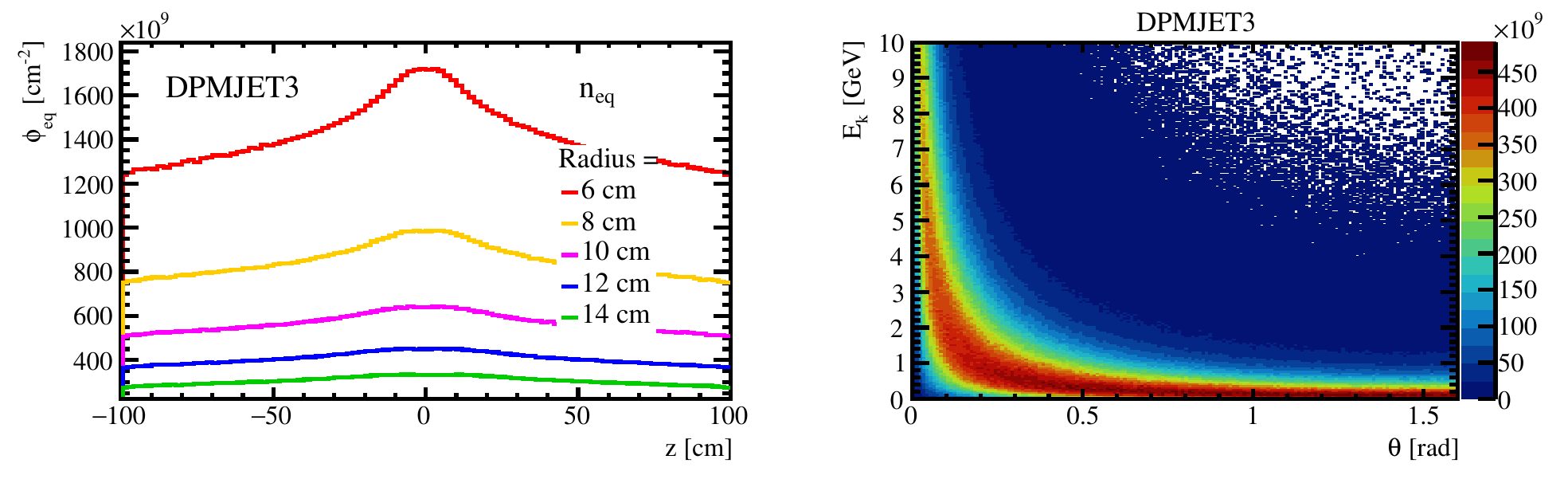} } 
\caption{Dependency of the neutron equivalence fluence simulated with DPMJET 3 on radius and distance from IP (left). The dependence of the kinetic energy on the polar angle (right). Values are scaled to 1 fb$^{-1}$ ($\sigma_{pp}$= 80 mb).} 
\label{fig:neq_z_ekin}
\end{figure}

The $z$-dependency of $\phi_{eq}$ for layers of detectors with different radii is presented in Fig.~\ref{fig:neq_z_ekin}. Particles traversing the detector in the central region have lower energies than those at very low polar angles (compare Fig.~\ref{fig:eta_en} and \ref{fig:neq_z_ekin}). Therefore, particles produced in at the highest angles, 
although they constitute less than 25\% of the total amount of particles, contribute the most significant to $\phi_{eq}$. The neutron equivalence fluence  peaks at the $z$=0 cm (contrary to the particle fluence) and saturates slower at higher distances than distributions of particles fluence. The decrease of $\phi_{eq}$ along the $z$ direction is due to particles with higher energy which are more often produced at very low polar angles and contribute to $\phi_{eq}$ with the lower value of the damage function $D(E)$. 

\subsection{Comparison of particle fluence}
\begin{figure}[htb]
\centerline{
\includegraphics[width=16cm]{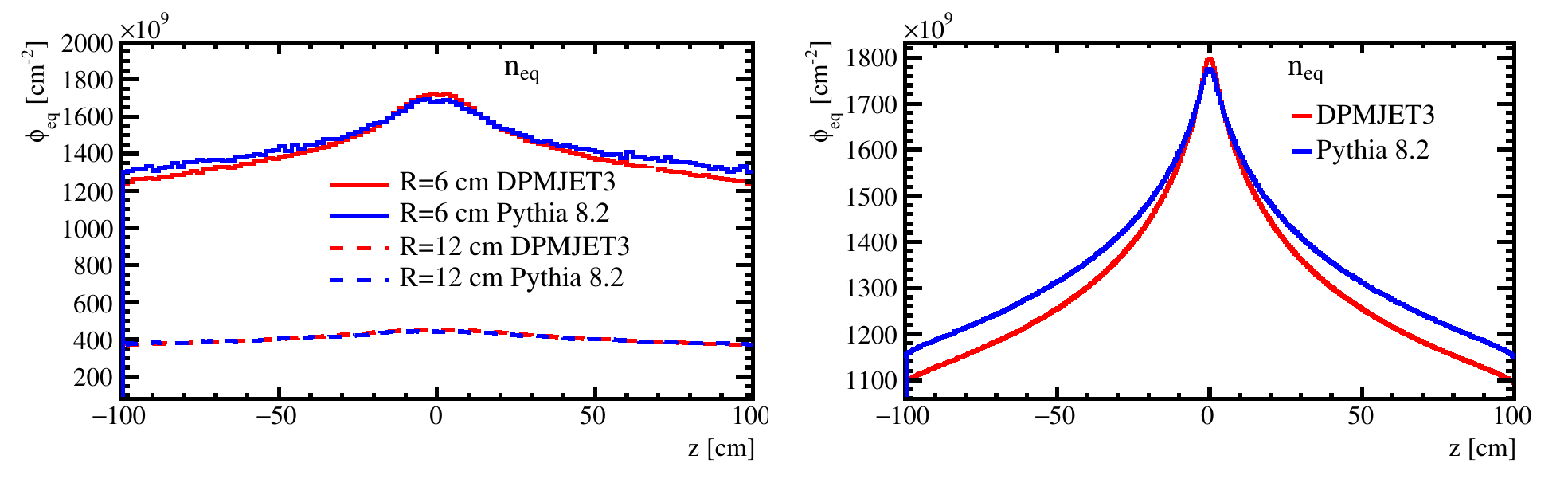} }
\caption{Comparison of DPMJET 3 and Pythia 8.2 simulation of neutron equivalence fluence as a function of $z$ (distance from IP along the beamline). Fluence is calculated at radius 6 cm (solid line) and 12 cm (dashed line) (left). Distribution of fluence averaged over radii 0.5 cm to 14 cm (right). Values are scaled to 1 fb$^{-1}$ ($\sigma_{pp}$= 80 mb).} 
\label{fig:neq_FP}
\end{figure}

Dependency of the neutron equivalent fluence $\phi_{eq}$ on the radius and the distance from the IP is shown in Fig.~\ref{fig:neq_FP} (left). At the radius $R$= 6 cm fluence reaches the value
$\phi_{eq}$=1.7$\times$10$^{12}$ n$_{eq}$/cm$^2$
for 1 fb$^{-1}$ of data, whereas the detectors at $z$=100 cm are 30\% less irradiated. This ratio drops to 16\% in case of stations at $R$=12 cm and decreases as radius increases. Fluence, averaged over the radius between 0.5 cm and 14 cm, is depicted in the right pane of Fig.~\ref{fig:neq_FP}. The lower radius is taken for the calculation, the greater the $z$-dependency in fluence is visible.

Comparing the DPMJET 3 and Pythia 8.2 simulations one can see agreement in detectors very close to the IP but in the more distant places, small excess (8\% at $R$=6 cm) of Pythia 8.2 fluence is noticeable. This can be explained while comparing the number and energy of pions produced in each generator. Although Pythia 8.2 generates up to 10\% (depending on the tuning) more particles than DPMJET 3 in one $pp$ collision, for the latter the fraction of generated pions with respect to all hadrons is higher. Therefore, the number of pions expected for the amount of data that corresponds to 1 fb$^{-1}$ of integrated luminosity in case of Pythia 8.2 is almost 10\% lower than in DPMJET 3. In the low-$z$ regions (high polar angle $\theta$) the highest contribution to $\phi_{eq}$ comes from pions; therefore DPMJET 3 shows small excess of $\phi_{eq}$ at $z$=0 cm in comparison with Pythia 8.2. 
In the case of protons, which populates small polar angles, Pythia 8.2 generates about 40\% more particles than DPMJET 3, which give higher fluence at a higher $z$. 

\begin{figure}[htb]
\centerline{
\includegraphics[width=16cm]{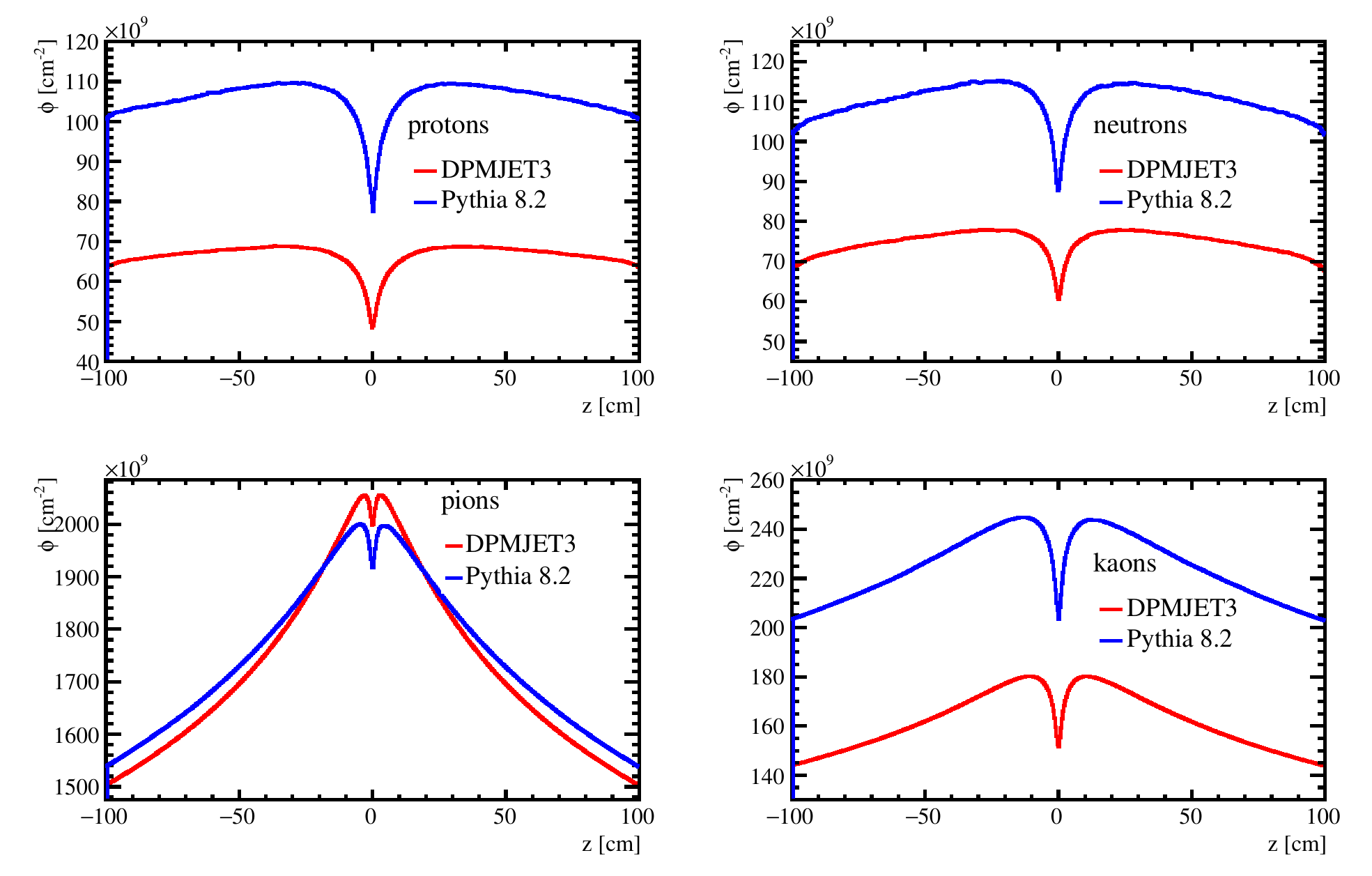} }
\caption{Fluence of protons, neutrons, pions and kaons simulated in FLUKA with DPMJET 3 and Pythia 8.2 generators as a function of the $z$-distance from IP. Fluence is averaged over radius from 0.5 cm to 14 cm. Values are scaled to 1 fb$^{-1}$ ($\sigma_{pp}$= 80 mb).} 
\label{fig:particle_fluence}
\end{figure}

The comparison of particle fluence for all types of hadrons simulated with the two generators is shown in Fig.~\ref{fig:particle_fluence}. It is visible that fluence of protons, neutrons and kaons diverges significantly among generators, but in case of pions, the difference is much smaller. The contribution of pions in the neutron equivalence fluence is the most substantial, therefore smaller difference between generators is caused mainly by the smaller difference in the pion component. In the more distant places, protons and neutrons contribute in larger amount and differences are more significant. 

Comparison of neutron equivalence fluence simulated by DPMJET 3 and Pythia 8.2 in various radius $R$ and $z$-positions is summarised in Table~\ref{tab:neq}.
\begin{table}[h]
\centering
\begin{tabular}{c|c|c|c|c|c|c}
\multirow{2}{*}{$\phi_{eq}$ [10$^{12}$/cm$^2$]} & \multicolumn{2}{c|}{$R$ = 0.5 cm} & \multicolumn{2}{c|}{$R$ = 6 cm} & \multicolumn{2}{c}{$R$ = 12 cm} \\
& D & P & D & P & D & P \\ \hline \hline
z = 0 cm & 216 & 214 & 1.71 & 1.66 & 0.45 & 0.44 \\
z = 100 cm& 80 & 86 & 1.22 & 1.32 & 0.36 & 0.37 
\end{tabular}
\caption{Neutron equivalent fluence $\phi_{eq}$ calculated in DPMJET 3 (D) and Pythia 8.2 (P) simulations at different radii ($R$) and distance from IP in $z$-direction. Fluence is calculated for 1 fb$^{-1}$ of data ($\sigma_{pp}$= 80 mb).}
\label{tab:neq}
\end{table}

The statistical uncertainties in case of Monte Carlo samples are negligible. The most significant contribution to systematic uncertainty comes from the damage functions parame\-tri\-sation used for the neutron equivalence calculation. According to the latest recommendations it may reach as much as 30\%~\cite{rd50}. 
In presented comparison this effect is negligible since the same (embedded in FLUKA) damage functions were applied for both generators.

\section{Summary}

The current and future High Energy experiments require precise reconstruction of tracks, vertices of production and decay in events that comprise hundreds of particles. Silicon trackers are usually situated in the densest radiation environment and are under the influence of the mixed radiation field. It has been proved that the currently used silicon structures should operate without compromising the quality of the physics data up to $\phi_{eq}$=10$^{16}$ n$_{eq}$/cm$^2$~\cite{rd50}. The fluence in the proximity to the IP, predicted for Run III at LHC, may reach this limit and will be well above it in the foreseen future experiments like HL-LHC~\cite{HL} and FCC (Future Circular Collider) \cite{FCC,FCC2}. Fluence can hardly be measured and must be rather simulated instead.
Therefore, it is vital to perform a detailed comparison of available tools for the generation of proton-proton events at LHC energies and transport codes used for the simulation of  track reconstruction.

Extensive studies of the most commonly used generators: Pythia 8.2 and DPMJET 3 showed that the former produces about 10\% more hadrons in each proton-proton collision. This difference originates from different physics model used to describe proton-proton interactions: Pythia generators use pQCD for the parton-parton interactions, DPMJET 3 model originates from soft hadronic interactions populated with hard pomeron exchange and multiple parton scattering for high-$p_T$ processes. Pythia models are tuned to the LHC data, which is predominantly obtained from the central detectors \cite{minbias_pred}. DPMJET 3 was revised after the LHC Run I to correct the problem of deficit of the high-multiplicity events \cite{FLUKA_rev}. There are no significant differences in the distributions of pseudorapidity, energy and transverse momentum in the central regions, but in DPMJET 3 more high energy particles in the very forward direction are generated. 

The fluence of particles in the inner part of detectors consists almost entirely of the primary particles produced in proton-proton interaction. Since pions constitute about 80\% of the total number of hadrons, the distribution of $\phi_{eq}$ is the most sensitive to the discrepancy in the production of pions. DPMJET 3 generates less particles per event but with a higher fraction of pions than Pythia 8.2, hence both generators show a reasonable agreement in the distribution of $\phi_{eq}$ for the regions very close to the IP. The influence of protons and neutron grows up for regions at high-$z$ tails. Therefore the difference in fluence simulated by Pythia 8.2 in a given $z$-position amounts to 7\% over the DPMJET 3. The $z$-dependency of the $\phi_{eq}$ becomes almost uniform for the more distant radial regions of the detector.

This analysis also showed the significant impact of Pythia generator parameters settings on the multiplicity distributions and the effect of minimum bias selection criteria on the number of particles that are further transported through the detector material to perform track recosntruction. Since DPMJET 3 generates more particles in the forward direction, a larger fraction of them is rejected by the selection cuts. This observation is important if $\phi_{eq}$ is simulated by methods that depend on the tracks reconstructed by standard algorithms since only tracks reconstructed with good quality and within the detector acceptance are further processed. It should also be considered while comparing the simulation among LHC experiments due to differences in the definition of minimum bias events.

\acknowledgments
This work was supported by the National Research Centre, Poland (NCN), grants No. UMO-2016/21/B/ST2/01083.


\end{document}